\documentclass[aps,12pt,showpacs,floatfix]{revtex4}
\usepackage{amsmath,amssymb,amsfonts,epsfig,bm}
\makeatletter
\renewcommand{\@makefntext}[1]{\parindent=1em\noindent\hbox to
1.8em{\hss$^{\@thefnmark}$}#1}
\renewcommand{\@footnotemark}{\hbox{\mathsurround=0pt$^{\@thefnmark}$}}
\newcommand{\ftnote}[2]{\footnotemark[#1]\footnotetext[#1]{#2}}
\makeatother
\newcommand{\be}{\begin{equation}}
\newcommand{\ee}{\end{equation}}
\newcommand{\ds}{\displaystyle}
\newcommand{\F}{Flatt{\'e}}

\begin{document}
 
\title{Nature of X(3872) from data}

\author{Yu. S. Kalashnikova}
\affiliation{Institute of Theoretical and Experimental Physics, 117218,
B.Cheremushkinskaya 25, Moscow, Russia} 
 
\author{A. V. Nefediev}
\affiliation{Institute of Theoretical and Experimental Physics, 117218,
B.Cheremushkinskaya 25, Moscow, Russia} 

\begin{abstract}
Properties of the charmonium-like state $X(3872)$ are investigated and its
nature is discussed as based on the existing experimental data. In particular,
we analyse the new data from Belle and BaBar Collaborations and argue that,
while the BaBar data prefer the dynamically generated virtual state in the $D\bar{D}^*$ system, the new Belle data clearly indicate a
sizable  $c \bar c$ $2^3P_1$ component in the $X$ wave function.

\end{abstract}
\maketitle

\section{Introduction}

In a few recent years a number of new states in the spectrum of charmonium have
been found experimentally. These states, labelled as $X$'s, $Y$'s, and $Z$'s
attract special attention of phenomenologists since most of them (if not all)
can hardly fit into the standard quark model scheme. This means that, in
addition to the genuine $c\bar{c}$ component, the wave functions of these
states must have extra components, whose nature is not yet clear
and is an open problem. Although various scenario are suggested and discussed
in the literature, such as threshold phenomena, hadronic molecule, and so on,
no unambiguous criteria which would allow one to distinguish between different
assignments for these ``homeless'' charmonia have been established so far. 

Among these new charmonium-like states the
$X(3872)$ meson is most well-studied. The $X$ was first observed in 2003 by the
Belle Collaboration in charged $B$-meson decays $B\to KX$, in the mode
$\pi^+\pi^-J/\psi$ \cite{Belle}, with the dipion originated from the
$\rho$-meson. 
The mass and the width of the $X(3872)$ reported in Ref.~\cite{Belle} were
\be
M_X=3872.0\pm 0.6(\mbox{stat})\pm
0.5(\mbox{syst})\mbox{MeV}
\ee
and
\be
\varGamma_X<2.3\mbox{MeV}.
\ee

Later Belle reported their observation of the same state in the
$\pi^+\pi^-\pi^0 J/\psi$ $(\omega J/\psi)$ and $\gamma J/\psi$ modes
\cite{Belleomega}, with branching fractions
\be
Br(B\to KX)Br(X\to\gamma J/\psi) = (1.8\pm 0.6\pm 0.1)\times 10^{-6},
\ee
\be
\frac{Br(X \to \pi^+\pi^-\pi^0J/\psi)}{Br(X \to
\pi^+\pi^-J/\psi)}=1.0 \pm 0.4 \pm 0.3,
\label{omega}
\ee
\be
\frac{Br(X \to \gamma J/\psi)}{Br(X \to
\pi^+\pi^-J/\psi)}=0.14 \pm 0.05.
\label{gamma}
\ee

The $X(3872)$ was confirmed in the discovery mode by the CDF
\cite{CDF}, D$\emptyset$ \cite{D0}, and BaBar \cite{BaBar} Collaborations.
In their recent updates, BaBar \cite{BaBar2} and Belle
\cite{Belle2} reduced slightly the branching ratio,
\be
Br(B^+\to K^+ X)\,Br(X\to \pi^+\pi^-
J/\psi)=(7\div 10)\times 10^{-6},
\label{BrA}
\ee
and both Belle and BaBar also observed the $X$ in the $B^0$ decays with
the rate comparable with the charged channel \cite{BaBar2, Belle2}.
In addition, BaBar measured other decay modes of the
$X$ \cite{BaBarrad}:
\be
\frac{Br(X\to \gamma
J/\psi)}{Br(X\to \pi^+\pi^- J/\psi)}=0.33\pm 0.12,
\label{raddec}
\ee
\be
\frac{Br(X\to \gamma \psi')}{Br(X\to \pi^+\pi^-
J/\psi)}=1.1\pm 0.4,
\label{raddecprime}
\ee
and imposed the upper limit on the $X$ production \cite{total}:
\be
Br(B \to K X)<3.2 \cdot 10^{-4}.
\label{uplim}
\ee

The most recent CDF result for the mass of the $X$
observed in the $\pi^+\pi^-J/\psi$ mode \cite{CDF3872} is
\be
M_X= 3871.61\pm 0.16\pm 0.19\,\mbox{MeV}\label{M:X3872-CDF}. 
\ee

Finally, the quantum numbers $J^{PC}=1^{++}$ are favoured for the $X$ (although
$2^{-+}$ are not yet excluded) \cite{XJPC}.

Clearly the measured properties of the $X(3872)$ raise a number questions
concerning its nature. In particular, although the $X$ is produced in the
$B$-mesons decays with the branching ratio of order $10^{-4}$ --- see
Eq.~(\ref{uplim}), that is with the branching ratio typical for genuine
charmonia (such as $J/\psi$, $\psi'$, or $\chi_{c1}$) \cite{PDG},
quark models fail to predict the existence
of a ${}^3P_1$ $c\bar{c}$ meson in the vicinity of the observed mass of 3872
MeV (see, for example, \cite{Hch}). In addition, quark--antiquark interpretation
of the
$X$ faces a further
challenge,
namely, a strong isospin violation --- see Eq.~(\ref{omega}).

In the meantime, the observed
mass of the $X(3872)$  measured in
the $\pi^+\pi^-J/\psi$ channel,
appears to be quite close to the position of the
$D^0\bar{D}^{*0}$ threshold which, according to the most recent CLEO data
\cite{dmass} lies at
\be
M_{D^0\bar{D}^{*0}}=3871.81\pm 0.36~\mbox{MeV}.
\label{MDDst}
\ee
It is quite natural to assume then that the $X$ wave function contains a large
admixture of the $D\bar{D}^*$ molecule component\ftnote{1}{An obvious shorthand
notation is
used here and in what follows: $D \bar D^* \equiv \frac{1}{\sqrt{2}}(D \bar
D^* + \bar D D^*)$.}, and the isospin violation is readily explained as due to
the large (about $7$ MeV) mass difference between the charged and neutral $D
\bar
D^*$ thresholds. Indeed, if one assumes that the decays $X \to \rho J/\psi$ and
$X \to \omega J/\psi$ proceed via $D \bar D^*$ loops, then the isospin violation
happens due to the difference between the charged and neutral loops which, in
turn, is due to the aforementioned mass difference.  Although it is not
large {\it per se}, it is enhanced due to kinematical reasons, as the effective
phase space available in case of the $\rho$ is much larger than that in case
of the $\omega$ \cite{suzuki,oset}.  

To summarise, data seem to indicate a dynamical origin of the
$X$. From the theoretical point of view several assignments for the latter are
discussed in the literature.

It was noticed long ago \cite{voloshin} that one-pion exchange can be
responsible for the formation of near--threshold
states in $D$--meson systems. In particular,
one-pion exchange is attractive in the $1^{++}$ $D \bar D^{*}$ channel
\cite{voloshin,tornqvist,swanson,others}. The $X(3872)$ as a virtual state
was discussed in Ref.~\cite{oset}, as generated dynamically from the interaction
of pseudoscalar and vector mesons.

On the other hand, in Ref.~\cite {YuSK}, in the framework of a coupled--channel
microscopic
quark model with the $c \bar c-D\bar{D}^*$ mixing, the $X(3872)$ is
generated as
a virtual state in the $D \bar D^*$ channel together with the $2^3P_1$
charmonium resonance. Similar phenomenon is found in a coupled--channel
analysis \cite{DS}, where a more sophisticated QCD-motivated approach to
light-quark pair creation is developed. 

The $X$ as a loosely $D \bar{D}^*$ bound state is
advocated in Refs.~\cite{braaten1,braaten2}, while the cusp scenario for the
$\pi^+ \pi^- J/\psi$
excitation curve in the $X(3872)$ mass range is discussed in Ref.~\cite{Bugg}.

Meanwhile, although the molecule assignment for the $X$ seems to be quite
plausible, it meets certain obstacles as well. To begin with,
a natural worry is that in the $D \bar D^*$ system, bound
by the one-pion exchange, the pion may go on-shell and thus binding may not be
strong enough \cite{suzuki}. For the most recent work on the possibility for
one-pion exchange to bind the $D \bar D^*$ system see \cite{ThCl,ch}.  Further
implications of the nearby pion threshold are discussed in
Refs.~\cite{braatenpions,pions}. 

Furthermore, quite a large
branching ratio for the radiative decay $X \to \gamma \psi'$ (see 
Eq.~(\ref{raddecprime})) can be explained naturally in the framework of quark
models.
Indeed, it is well-known that in so-called Coulomb+linear quark potential
models the radiative decay $\chi'_{c1} \to \gamma J/\psi$ is suppressed in
comparison with decay $\chi'_{c1} \to \gamma \psi'$. For example, the
estimates of Refs.~\cite{Hch} and \cite{BG} yield:
\begin{eqnarray}
\varGamma(\chi'_{c1}\to\gamma J/\psi)=70~\mbox{keV~\cite{Hch}},\quad
11~\mbox{keV~\cite{BG}},\nonumber\\[-2mm]
\label{radquark}\\[-2mm]
\varGamma(\chi'_{c1}\to\gamma\psi')=180~\mbox{keV~\cite{Hch}},\quad
64~\mbox{keV~\cite{BG}}.
\nonumber
\end{eqnarray}
In the molecular model an opposite pattern was found in Ref.~\cite{swanson}, and
a
large $\gamma \psi'$ rate is now  considered as an evidence against the molecule
interpretation. 
Notice, however, that there exists a mechanism for the radiative decays of
molecules via $D^{(*)}$--meson loops which was not considered in
Ref.~\cite{swanson}
and which favours the $\chi'_{c1} \to \gamma \psi'$ decay rate
over the $\chi'_{c1} \to \gamma J/\psi$ one. However, a reliable evaluation of
such
radiative decays of molecules meets severe problems with divergent loop
integrals which can hardly be resolved in a model-independent way.
 
Finally, for a pure molecule, the branching fraction $B \to K X$ was estimated
in Ref.~\cite{braatenmolrate} to be less than $10^{-5}$, that is much smaller
than
the experimental data on the $X$ production (though, being very model-dependent,
such estimates should be treated with caution). So it seems quite
reasonable to assume that this is the $c \bar c$ component of the $X$
to be responsible for the $X$ production in $B$ meson decays and for the $X$
radiative decays.

The interest to the $X(3872)$ was catalysed even more in 2006, when Belle
reported
an enhancement of the $D^0\bar D^0\pi^0$ signal just above the $D^0\bar
{D}^{*0}$ threshold observed in the
reaction $B^+\to K^+D^0\bar{D}^0\pi^0$ \cite{Belletalk,Belle3875} at
\be
M_X=3875.2\pm 0.7^{+0.3}_{-1.6}\pm 0.8~\mbox{MeV},
\label{3875v1}
\ee
with the branching
\be 
Br(B^+\to K^+D^0\bar{D}^0\pi^0)=(1.02\pm
0.31^{+0.21}_{-0.29})\times 10^{-4}.
\label{BrB}
\ee
The corresponding state was called the $X(3875)$ and it was confirmed later by
the BaBar Collaboration as well \cite{Babarddpi}.

Although an immediate and the most natural conclusion is that this is simply
yet another manifestation of the same well-established state $X(3872)$, a 3~MeV
shift in the mass may have had dramatic consequences for such an interpretation.
As a result, a rather extreme assumption was made that two different
charmonium-like states might reside in the same mass region.
It was noticed in Ref.~\cite{recon}, however, that, under certain assumptions on
the
nature of the $X$, the two states could be indeed reconciled with one another.
In particular, it was argued in Ref.~\cite{recon} that the $X$, as a virtual
state
in the $D\bar{D}^*$ system, can reproduce both sets of data, for the $DD\pi$
and $\rho J/\psi$ channels, whereas in the latter case one deals with a
threshold cusp. Parameters of the model were tuned to fit all the data on
the resonance width and branching ratios. These results of Ref.~\cite{recon}
appear
to be in a good agreement with the findings of Refs.~\cite{YuSK,Bugg}. 
The analysis
of Ref.~\cite{recon} is improved in Ref.~\cite{YuSKconf} where additional non--$D
\bar D^{*}$ modes of
the $X$ were taken into account and the admixture of the genuine $c\bar{c}$
charmonium in the $X$ wave function was estimated. 

Recently Belle Collaboration announced a new analysis for the 
$D^{*0}\bar{D}^0$ case \cite{Belle3875v2}. The new data on the
$D^{*0}\to D^0\pi^0$ and $D^{*0}\to D^0\gamma$ channels were fitted
both with the simple Breit--Wigner line--shape form and with the \F~formula. 
As a result, a lower position of the peak,
\be
M_X= 3872.6^{+0.5}_{-0.4}\pm 0.4~\mbox{MeV},
\ee
was obtained than the one reported before --- see Eq.~(\ref{3875v1}). The
corresponding branching ratio was measured to be
\be 
Br(B^+\to
K^+X(D^{*0}\bar D^0))=(0.73\pm 0.17\pm 0.13)\times 10^{-4}.
\ee
These new data are analysed in Ref.~\cite{ch2} using the technique very close to
that of Ref.~\cite{recon}, and the conclusion is made 
that the $X$ is a 2$^3P_1$ $\bar cc$ state strongly distorted by couple channel effects.

In this paper we present an updated \F~analysis, with the new data on the $D^0
\bar D^{*0}$ mode \cite{Belle3875v2} and on the $\gamma \psi'$ mode
\cite{BaBarrad}
included. In particular, we address the question of a possible $\chi'_{c1}$
charmonium admixture in the wave function of the $X(3872)$. The strategy
employed in this paper, differs significantly from the one of Ref.~\cite{recon},
where a model-blind \F~analysis was performed. Here we assume a mechanism for
the $X$ production via the charmonium component.

The paper is organised as follows. In Section~\ref{flattesect} we give
necessary details of the \F~parametrisation for a near-threshold
resonance and apply this technique to the case of the $X(3872)$. We analyse the data in Section~\ref{datasect}, and comment on the effect of the $D^{*0}$ finite width in Section~\ref{finwidth}.  We conclude and discuss the results in
Section~\ref{discsect}.

\section{\F~parametrisation}\label{flattesect}

In this Section we introduce a \F-like
parametrisation of the near--threshold observables related to the $X(3872)$
state.
Let us define the energy $E$ relative to the neutral $D^0\bar{D}^{*0}$
threshold (see Eq.~(\ref{MDDst})). Then the relevant energy range is
approximately $-10\;\mbox{MeV}\lesssim E\lesssim 10\;\mbox{MeV}$ which covers
both the three-body $D^0 \bar D^0\pi^0$ threshold at $E_{D^0 \bar D^0
\pi^0}\approx -7$ MeV and the charged
$D^+\bar{D}^{*-}$ threshold at $E_{D^+\bar{D}^{*-}}\equiv\delta\approx 7.6$
MeV. 
A natural generalisation of the standard \F~parametrisation for the
near--threshold
resonance \cite{Flatte} of the $D^0 \bar D^{*0}$
scattering amplitude reads
\be
F(E)=-\frac{1}{2k_1}\frac{g_1k_1}{D(E)},
\label{flatte2}
\ee
with
\be
D(E)=\left\{
\begin{array}{ll}
\ds
E-E_f-\frac{g_1\kappa_1}{2}-\frac{g_2\kappa_2}{2}+i\frac{\varGamma(E)}{2},&E<0\\
[3mm]
\ds
E-E_f-\frac{g_2\kappa_2}{2}+i\left(\frac{g_1k_1}{2}+\frac{\varGamma(E)}{2}
\right),&0<E<\delta\\[3mm]
\ds
E-E_f+i\left(\frac{g_1k_1}{2}+\frac{g_2k_2}{2}+\frac{\varGamma(E)}{2}\right),
&E>\delta
\end{array}
\right.
\label{D}
\ee
and
$$
k_1=\sqrt{2\mu_1 E},\quad\kappa_1=\sqrt{-2\mu_1 E},\quad
k_2=\sqrt{2\mu_2(E-\delta)},\quad\kappa_2=\sqrt{2\mu_2(\delta-E)}.
$$
Here $\mu_1$ and $\mu_2$ are the reduced masses in the $D^0
\bar D^{*0}$ and $D^+ D^{*-}$ channels, respectively. Assuming 
isospin conservation we set $g_1=g_2=g$.

The term $i\varGamma(E)/2$ in Eq.~(\ref{D}) accounts for
non-$D \bar D^*$ modes:
\be 
\varGamma(E)=\varGamma_{\pi^+\pi^-J/\psi}(E)+
\varGamma_{\pi^+\pi^-\pi^0J/\psi}(E)+\varGamma_0,
\ee 
where we single out the first two modes because of their
explicit energy dependence: 
\be
\varGamma_{\pi^+\pi^-J/\psi}(E)=f_{\rho}\int^{M-m_{J/\psi}}_{2m_{\pi}}
\frac{dm}{2\pi}\frac{q(m)\varGamma_{\rho}}{(m-m_{\rho})^2+\varGamma_{\rho}^2/4},
\label{rhowidth}
\ee
\be
\varGamma_{\pi^+\pi^-\pi^0J/\psi}(E)=f_{\omega}\int^{M-m_{J/\psi}}_{3m_{\pi}}
\frac{dm}{2\pi}\frac{q(m)\varGamma_{\omega}}{(m-m_{\omega})^2+\varGamma_{\omega}
^2/4 } ,
\label{omegawidth}
\ee
with $f_{\rho}$ and $f_{\omega}$ being effective couplings and
\be
q(m)=\sqrt{\frac{(M^2-(m+m_{J/\psi})^2)(M^2-(m-m_{J/\psi})^2)}{4M^2}}
\ee
being the centre-of-mass dipion/tripion momentum ($M=E+M(D^0 \bar D^{*0})$).

Now, if we assume the short-ranged dynamics of the weak $B \to K$
transition to be absorbed into the coefficient ${\cal B}$,
then the differential rates of interest in the \F~approximation read:
\be
\frac{dBr(B \to K D^0 \bar D^{*0})}{dE}={\cal
B}\frac{1}{2\pi}\frac{gk_1}{|D(E)|^2},
\label{DD}
\ee
\be
\frac{dBr(B \to K \pi^+\pi^- J/\psi)}{dE}={\cal
B}\frac{1}{2\pi}\frac{\varGamma_{\pi^+\pi^-J/\psi}(E)}{|D(E)|^2},
\label{rhopsi}
\ee
and
\be
\frac{dBr(B \to K \pi^+\pi^-\pi^0 J/\psi)}{dE}={\cal
B}\frac{1}{2\pi}\frac{\varGamma_{\pi^+\pi^-\pi^0J/\psi}(E)}{|D(E)|^2}.
\label{omegapsi}
\ee

Obviously, the rate (\ref{DD}) is defined for $E>0$ only, while
the rates (\ref{rhopsi}) and (\ref{omegapsi}) are defined
both above and below the $D^0 \bar D^{*0}$ threshold. Strictly speaking, one is
to take into account a finite width of the $D^*$, which is very small however.
Indeed, the total width of the $D^{*\pm}$-meson is measured to be $96 \pm 22$
keV \cite{PDG}. There are no data on the $D^{*0}$ width, but one can estimate the total width of the $D^{*0}$ from the data \cite{PDG} on charged
$D^{*\pm}$ to be about $63$ keV, which gives $\varGamma(D^{*0} \to D^0 \pi^0)=42$ keV. If,
nevertheless, the finite width of the $D^*$ is taken into account, the
rate (\ref{DD}) continues to the region $E<0$ and interference effects
are possible in the final state, as described in Ref.~\cite{voloshinint}.
We shall discuss this in some detail below
though, at the moment, we follow Refs.~\cite{recon,YuSKconf} and neglect the
$D^*$
width. 

Now, in order to proceed to the branching ratio to the $DD\pi$ final state, one
is to take into account the branching fractions of the $D^{*0}$ \cite{PDG}:
\be
Br(D^{*0} \to D^0 \pi^0)=(61.9 \pm 2.9)\%,
\label{062}
\ee
\be
Br(D^{*0} \to D^0 \gamma) =(38.1 \pm 2.9)\%,
\label{038}
\ee 
so that
\be
\frac{dBr(B\to K D^0 \bar D^0 \pi^0)}{dE}=0.62{\cal B}
\frac{1}{2\pi}\frac{gk_1}{|D(E)|^2}.
\label{DDpi}
\ee
Analogously we have for the $D^0 \bar D^0 \gamma$ differential
rate:
\be
\frac{dBr(B\to K D^0 \bar D^0 \gamma)}{dE}=0.38{\cal B}
\frac{1}{2\pi}\frac{gk_1}{|D(E)|^2}.
\label{DDgamma}
\ee

With the \F~parametrisation introduced above, one can
make use of the method
suggested in Ref.~\cite{evi} to estimate the admixture of a bare $\chi'_{c1}$
state
in the wave function of the $X$. 
Indeed, in the context of $c \bar c$--$D \bar D^*$ coupled--channel model the
quantities entering the \F-type expressions for differential rates acquire
clear physical meaning. Namely, the coefficient ${\cal B}$ can be viewed as
the branching fraction $B \to K \chi'_{c1}$, $g$ is the bare $\chi'_{c1} D \bar
D^*$
coupling constant, and $\varGamma_0$ is the bare total width of the
$\chi'_{c1}$ level. 
Moreover, as shown in Ref.~\cite{evi}, in the \F~limit, the
probability $w(E)$
to find the bare state in the wave function of a physical state can be expressed in terms of \F~parameters as
\be
w(E)=\frac{1}{2\pi|D(E)|^2}(gk_1\Theta(E)+gk_2\Theta(E-\delta)+\varGamma(E)).
\label{wE}
\ee
The admixture $W$ of the $\chi'_{c1}$ charmonium in the resonance wave function can be defined as
\be
W=\int_{E_{\rm min}}^{E_{\rm max}} w(E)dE,
\label{Wint}
\ee
where the integral it taken over the near-threshold region. As was discussed
before, we choose is to be
from -10 MeV to +10 MeV.

\section{Data analysis}\label{datasect}

\subsection{Essentials and constraints}

In this chapter we analyse the existing data using the \F~approach
described above. Our aim is to estimate the admixture of the charmonium
component of the $X$ wave function and to identify the nature of the residual,
dynamically generated part of its wave function. In particular, we shall answer
the question as to whether the existing experimental data are compatible with
the bound state or virtual state in the $D^0D^{*0}$ system.

Let us comment briefly on the difference between the bound/virtual state
situations.

In the \F~approximation
the inelastic differential rate is 
\be
{\cal B}\frac{1}{2\pi}\frac{\varGamma(E)}{|D(E)|^2}.
\label{inelrate}
\ee
One can see easily that the behaviour of the
inelastic rate below threshold depends strongly on whether there is a zero in
the real part of the denominator $D(E)$ below threshold. Indeed, if $D(E_{\rm
bound})=0$ for some $E_{\rm bound}<0$ then, for $\varGamma(E) \to 0$,  the
inelastic rate
(\ref{inelrate}) does not vanish, but becomes proportional to
a $\delta$-function:
\be
{\cal B}\left(\frac{\partial D(E)}{\partial E}_{|_{E=E_B}}\right)^{-1} \delta
(E-E_{\rm bound}).
\label{delfunc}
\ee
We end up therefore with a real bound state, which is
not coupled to inelastic channels. 

On the contrary, if there is no such zero (virtual state case), the rate
(\ref{inelrate}) vanishes as $\varGamma(E)\to 0$, while the $D^0 \bar
D^{*0}$ rate does not vanish in this limit. 

Consider the case of $\varGamma_0=0$ first.
In this case, in order to distinguish between these two scenarios (bound state
versus virtual state) one is, as was argued in \cite{braaten2,recon}, to check the ratio 
\be
\frac{Br(X \to D^0 \bar D^0 \pi^0)}{Br(X \to \pi^+\pi^- J/\psi)},
\label{Sdpsi}
\ee
which varies from quite small (and hardly resolvable experimentally against the
background) values, for the bound state scenario, up to values of order ten
(in Ref.~\cite{recon} this ratio was calculated to be 9.9) for
the virtual state. It follows from the data quoted in Eqs.~(\ref{BrA}) and
(\ref{BrB}) that
the ratio (\ref{Sdpsi}) is indeed large ($\simeq 10\div 15$), which seems to
indicate the virtual state nature of the $X$. However, the above consideration
was based on the assumption that, once
produced, the $X$ state can only decay through one of the three channels:
$D^0\bar{D}^0\pi$ or $\rho J/\psi$ and $\omega J/\psi$ (the $\gamma J/\psi$ mode
is small, and was neglected). Nowadays, a new $\gamma \psi'$ mode is observed.
Moreover, if it is presumably due to $c \bar c$ bare seed, then extra decay
channels
typical for charmonium should exist for the $X$, which are encoded in the extra
width $\varGamma_0\neq 0$. These are annihilation
modes (into light hadrons), and $\chi_{c1}(3515) \pi\pi$ (the latter was
estimated in Ref.~\cite{voloshinpipi} to be of order of a few keV). The total
width of
the $\chi_{c1}(3515)$ is 0.89 $\pm$ 0.05 MeV, and the branching fraction into
radiative $\gamma J/\psi$ mode is about 36\% \cite{PDG}. If it were a true
guide, then one expects the width of the
$\chi'_{c1}$ to be about $1\div 2$ MeV. Quark model prediction \cite{BG} yields
the value of $1.72$ MeV for the total width of the $\chi'_{c1}$. In accordance
with predictions (\ref{radquark}), radiative modes are not the dominant ones.

Now the ratio (\ref{Sdpsi}) should be modified to read:
\be
\frac{Br(X \to D^0 \bar D^0 \pi^0)}{Br(X \to {\rm non}D^0 \bar D^0 \pi^0)}\sim 1,
\label{Sdpsi2}
\ee
opening the possibility for the $X$ being a bound state\ftnote{2}{The idea that,
including an extra width, one can fit the data on the $X(3872)$ both with
virtual and bound state was first presented in Ref.~\cite{bugg2}.}.

Finally, assuming the $X$ to be produced via the $\chi'_{c1}$ component of its
wave function, one can estimate the coefficient ${\cal B}$. The world average
for the $Br(B^+ \to K^+ \chi_{c1})$ is \cite{PDG}
\be
Br(B^+ \to K^+ \chi_{c1})=(5.1 \pm 0.5) \times 10^{-4},
\label{chi}
\ee
and it is known \cite{PDG} that $J/\psi$ and $\psi'$ are produced in the $B \to K$ decays
with comparable branching fractions:
\begin{eqnarray*}
Br(B^+ \to K^+ J/\psi)&=&(10.22 \pm 0.35) \times 10^{-4},\\
Br(B^+ \to K^+ \psi')&=&(6.48 \pm 0.35) \times 10^{-4}.
\end{eqnarray*}
Then it is reasonable to assume that
the $\chi'_{c1}$ is produced in the $B \to K$ decays with the rate comparable to
(\ref{chi}). There exists a quark model prediction \cite{chinesechi}
$Br(B \to K \chi'_{c1})=2 \times 10^{-4}$. However, the model used in
Ref.~\cite{chinesechi} underestimates the rate (\ref{chi}) more than two times.

As was mentioned before, the admixture of the genuine charmonium in the $X$
wave function is given by the quantity $W$ defined in Eqs.~(\ref{wE}) and
(\ref{Wint}).

Therefore, our analysis strategy is to approximate the existing experimental
data on the $D\bar{D}^*$ and $\pi^+\pi^- J/\psi$ decay modes of the $X$ with the
\F~formulae and
\begin{itemize}
\item to find the admixture of the $\chi_{1c}'$ charmonium in the $X$ wave
function by evaluating the integral (\ref{Wint}) of the spectral density
(\ref{wE}) over the near-threshold region;
\item to compute the scattering length for the $D\bar{D}^*$ system and thus to
make a conclusion concerning its virtual/bound state nature;
\item to investigate the effect of the finite width $\varGamma_0$.
\end{itemize}

The data on the $D\bar{D}^*$ and
$\pi^+\pi^-J/\psi$ modes are analysed under the following constraints:
\begin{itemize}
\item $Br(B \to K\chi'_{c1})={\cal B}=(3 \div 6) \cdot 10^{-4}$, with the
preference to lower values (see Eq.~(\ref{chi}) and the discussion following
it);
\item $Br(B \to K X)={\cal B}W<3.2 \cdot 10^{-4}$ (the limit imposed by the
BaBar data \cite{total}, see Eq.~(\ref{uplim}));
\item $\varGamma_0=1 \div 2$ MeV (as discussed above).
\end{itemize}

Throughout this paper we deal only with the data on the charged $B$--meson
decays, as the
uncertainties in the data on the neutral mode remain large. Belle Collaboration
presents the data on the $D^0 \bar D^0 \pi^0$ and $D^0 \bar D^0 \gamma$ modes
separately, and we analyse only the former mode (again due to larger
uncertainties in the $D^0 \bar D^0 \gamma$ mode). BaBar data presented are for
all $D^0 \bar D^{*0}$ modes, so we consider these data.

\subsection{Belle Collaboration data}

As it was mentioned in the introductory part, recently Belle Collaboration
presented a new analysis for the $\pi^+\pi^-J/\psi$, $D\bar{D}\pi$, and
$D\bar{D}\gamma$ decay modes of the $X$ \cite{Belle2,Belle3875v2}. These new
data differ significantly from the old ones. The peak in the $\pi^+\pi^-J/\psi$
mass distribution is shifted to the left, making the virtual state/cusp scenario
advocated in Ref.~\cite{recon} less plausible. However, as the ratio
(\ref{Sdpsi})
remains large, extra non--$D\bar D \pi$ modes are needed in order to arrive at
the bound--state solution, as it follows from Eq.~(\ref{Sdpsi2}) and will
be shown below.

\begin{table}[t]
\caption{The sets of the \protect{\F} parameters for the
Belle data from Refs.~\cite{Belle2,Belle3875v2}.}
\begin{ruledtabular}
\begin{tabular}{ccccccccccc}
Set&$\varGamma_0$&$g$&$E_f$, MeV&$f_\rho$&$f_\omega$&
${\cal B}\cdot 10^4$&$\phi$&$W$&${\cal B}W\cdot 10^4$&$a$, fm\\
\hline
1&1.1&0.3&-12.8&0.00770&0.04070&2.7&$180^0$&0.19&0.5&$-5.0-i1.3$\\
2&1.0&0.137&-12.3&0.00047&0.00271&4.3&$153^0$&0.43&1.9&$3.5-i1.0$\\
3&2.0&0.091&-7.8&0.00090&0.00523&3.7&$152^0$&0.52&1.9&$3.3-i1.7$\\
\end{tabular}
\end{ruledtabular}
\label{bellesets}
\end{table}

In order to translate the differential rates into number-of-events
distributions,
we notice that there are $131$ signal events in the Belle data for the $\pi^+
\pi^- J/\psi$ channel \cite{Belle2}, which corresponds to the branching fraction
of about $8.1 \cdot 10^{-6}$; the bin size is 2.5~MeV. Then
\be
N^{\pi \pi J/\psi}_{\rm Belle}(E)=2.5 \,  {\rm [MeV]} \left(\frac{131}{8.3 \cdot
10^{-6}}\right)
\frac{dBr(B \to K \pi^+ \pi^- J/\psi)}{dE} .
\label{rhonoebelle}
\ee

Similarly, for the $D^0 \bar D^0 \pi^0$ mode, the Belle Collaboration
states to have $48.3$ signal events in the charged mode
\cite{Belle3875v2}, which corresponds to the
branching fraction of about
$0.73 \cdot 10^{-4}$; the bin size is 2~MeV. Thus the
number-of-events distributions 
is calculated as
\be
N^{D^0 \bar D^0 \pi^0}_{\rm Belle}(E)=2.0 {\rm [MeV]} \left(\frac{48.3}{0.73 \cdot
10^{-4}}\right)
\frac{dBr(B \to K D^0 \bar D^{0}\pi^0)}{dE}.
\label{dnoe}
\ee

In the latter case, the background function
is proportional to the two--body $D^0 \bar D^{*0}$ phase space
$R_2\propto\sqrt{E}$, that is
the background is considered to be due to
the contribution of the $D^0 \bar D^{*0}$ and, as such, to interfere with the
signal:
\be
\frac{dBr(B \to
K D^0 \bar D^{0}\pi^0)}{dE}= 0.62\frac{k_1}{2\pi}\left[\left({\rm
Re}\frac{\sqrt{g \cal B}}{D(E)}
+c\cos\phi\right)^2 +\left({\rm Im}\frac{\sqrt{g \cal
B}}{D(E)}+c\sin\phi\right)^2\right],
\label{tot2}
\ee
with the relative phase $\phi$ and $c$ being fitting constants.

Finally, the resolution functions for both reactions are taken in the form of
Gaussians with the fixed resolution scale being 3~MeV, for the $\pi^+\pi^-
J/\psi$ channel, and with the variable mass-dependent resolution
function $\sigma(m)=a\sqrt{m-m_0}$, with $a=0.172$~MeV$^{1/2}$ and
$m_0=M(D\bar{D}^*)$ \cite{Belle3875v2}.

\begin{figure}
\epsfig{file=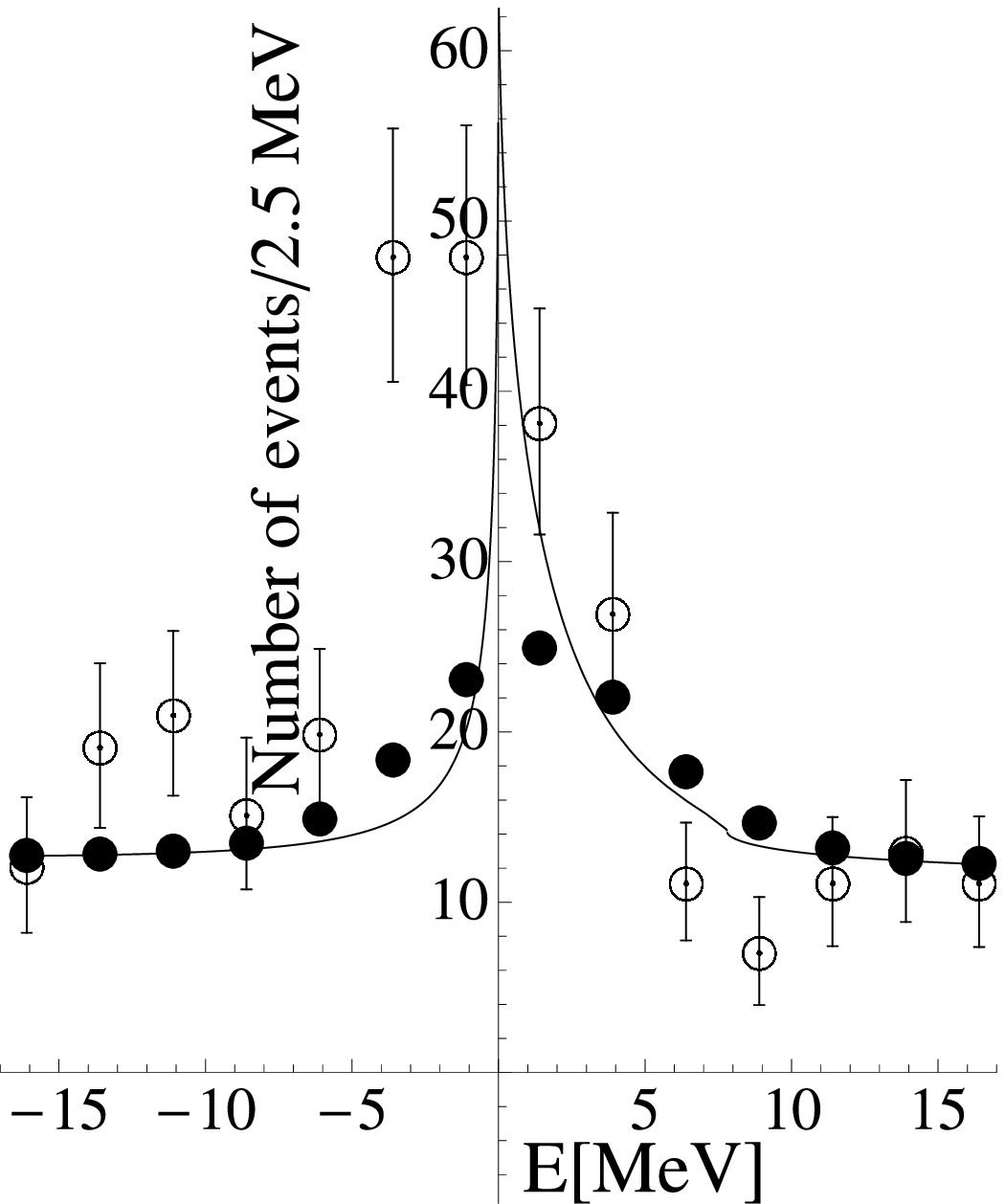,width=6cm}\hspace*{1cm}
\epsfig{file=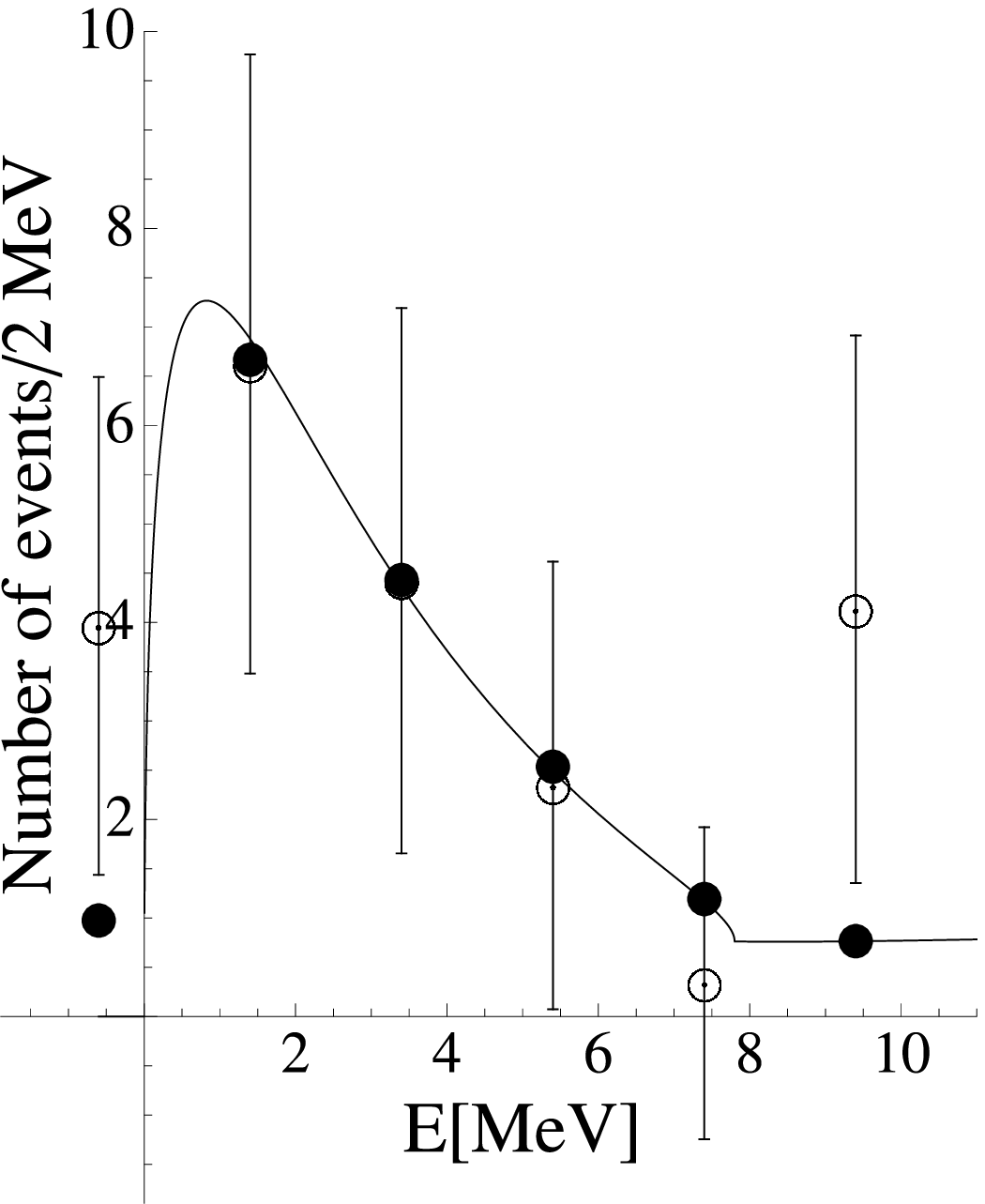,width=6cm}
\caption{Differential rates for the $\pi^+ \pi^-J/\psi$ channel (left plot) and
$D^0 \bar{D}^{0}\pi^0$ channel (right plot) (see Eqs.~(\ref{rhonoebelle}) and
(\ref{dnoe}), respectively) with the parameters given by set~1 (see
Table~\ref{bellesets}). The distributions integrated over the bins, with 
resolution function taken into account, are shown
as filled dots, experimental data (see Refs.~\cite{Belle2,Belle3875v2}) are
given as open dots with
error bars.}\label{virboundfig}
\end{figure}

\begin{figure}
\begin{center}
\begin{tabular}{ccc}
\epsfig{file=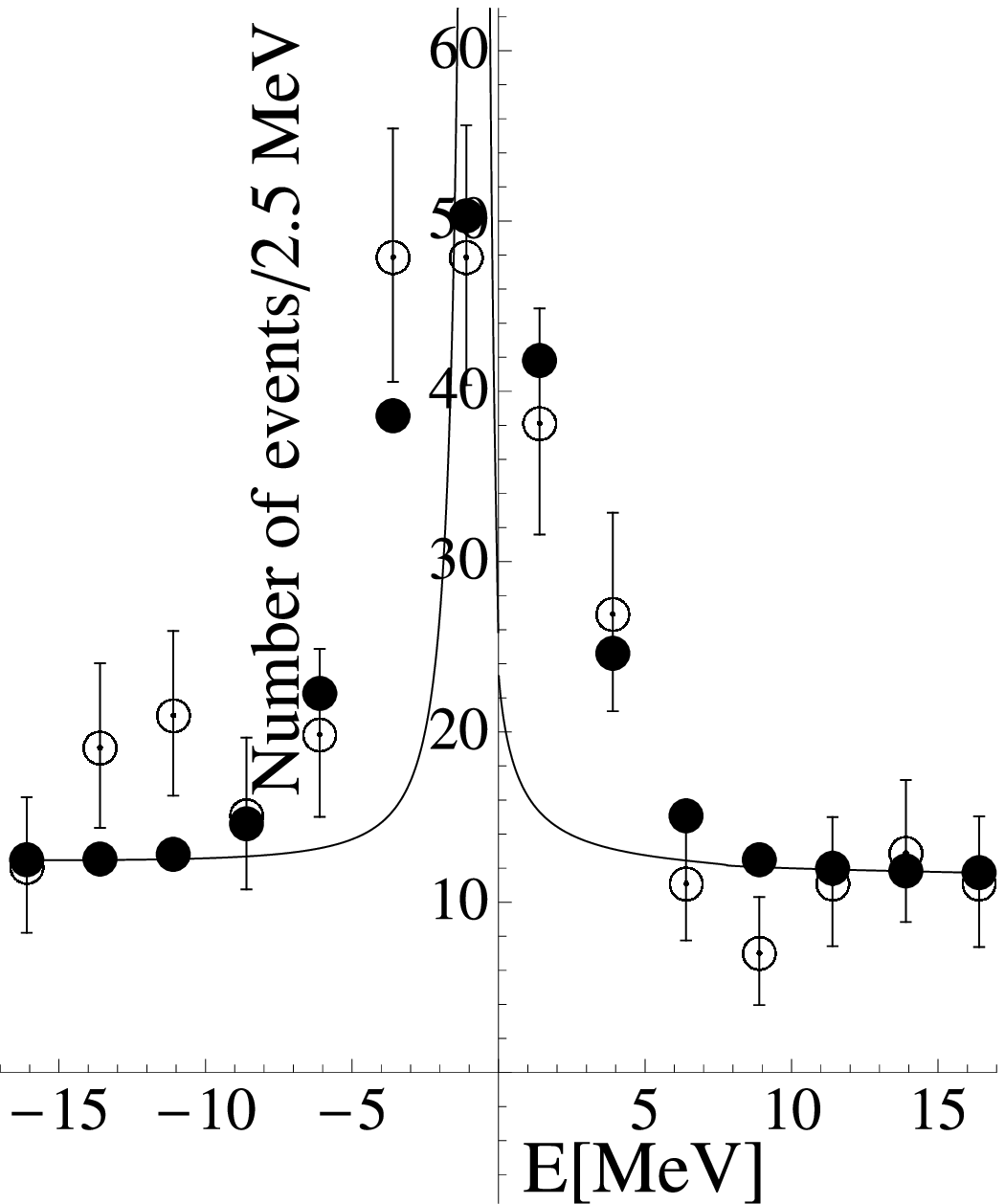,width=6cm}&\hspace*{1cm}&
\epsfig{file=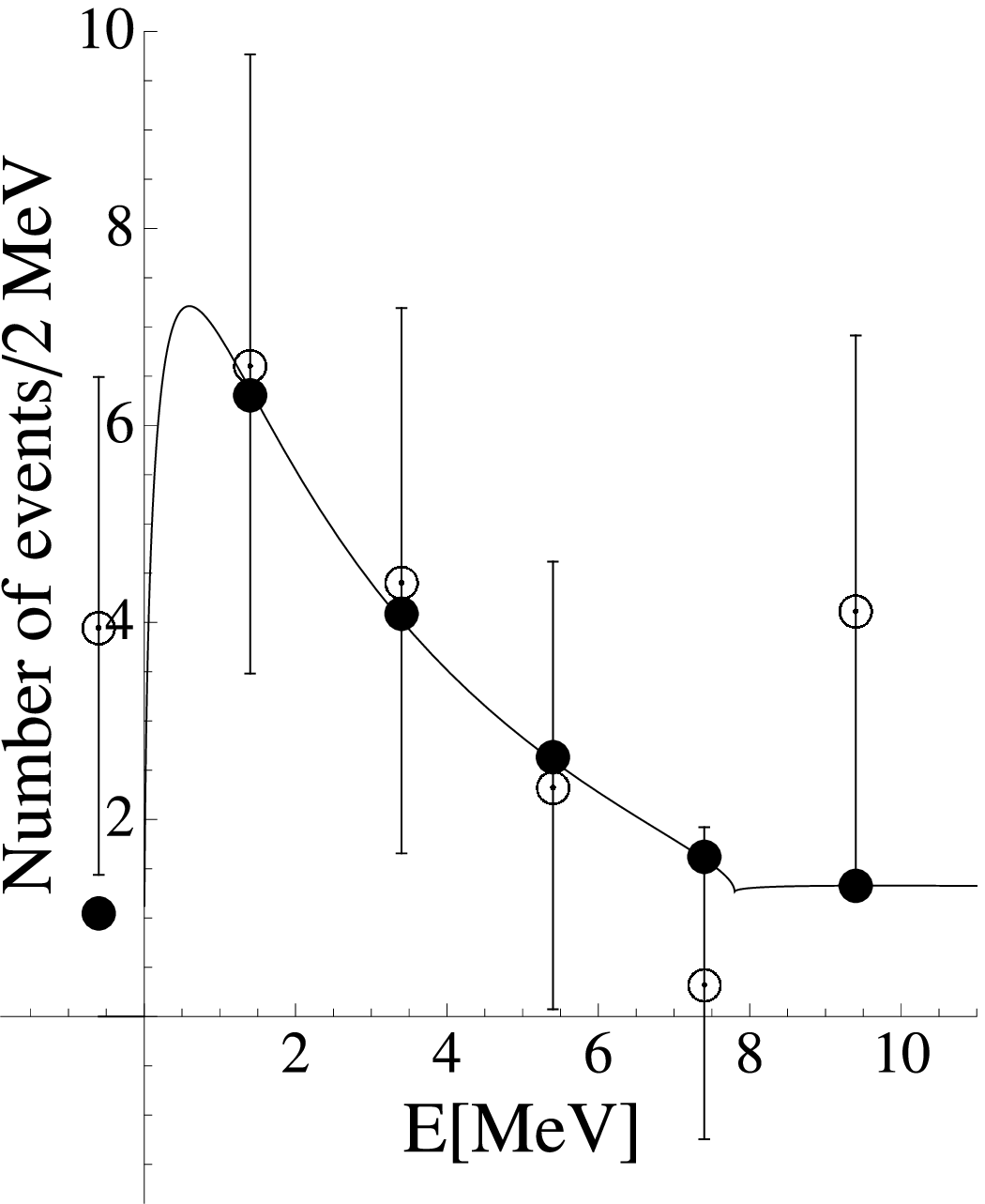,width=6cm}\\
\epsfig{file=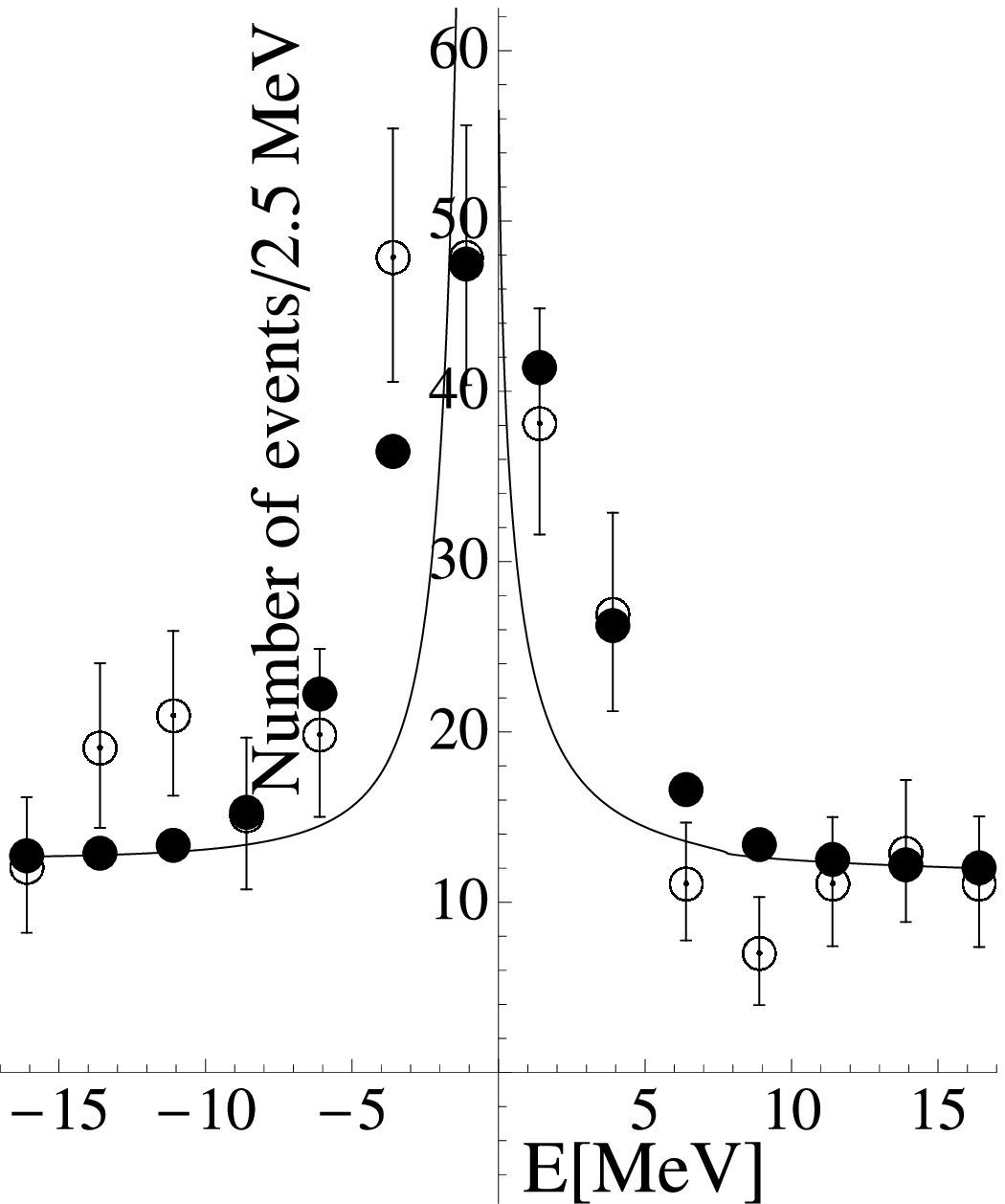,width=6cm}&&
\epsfig{file=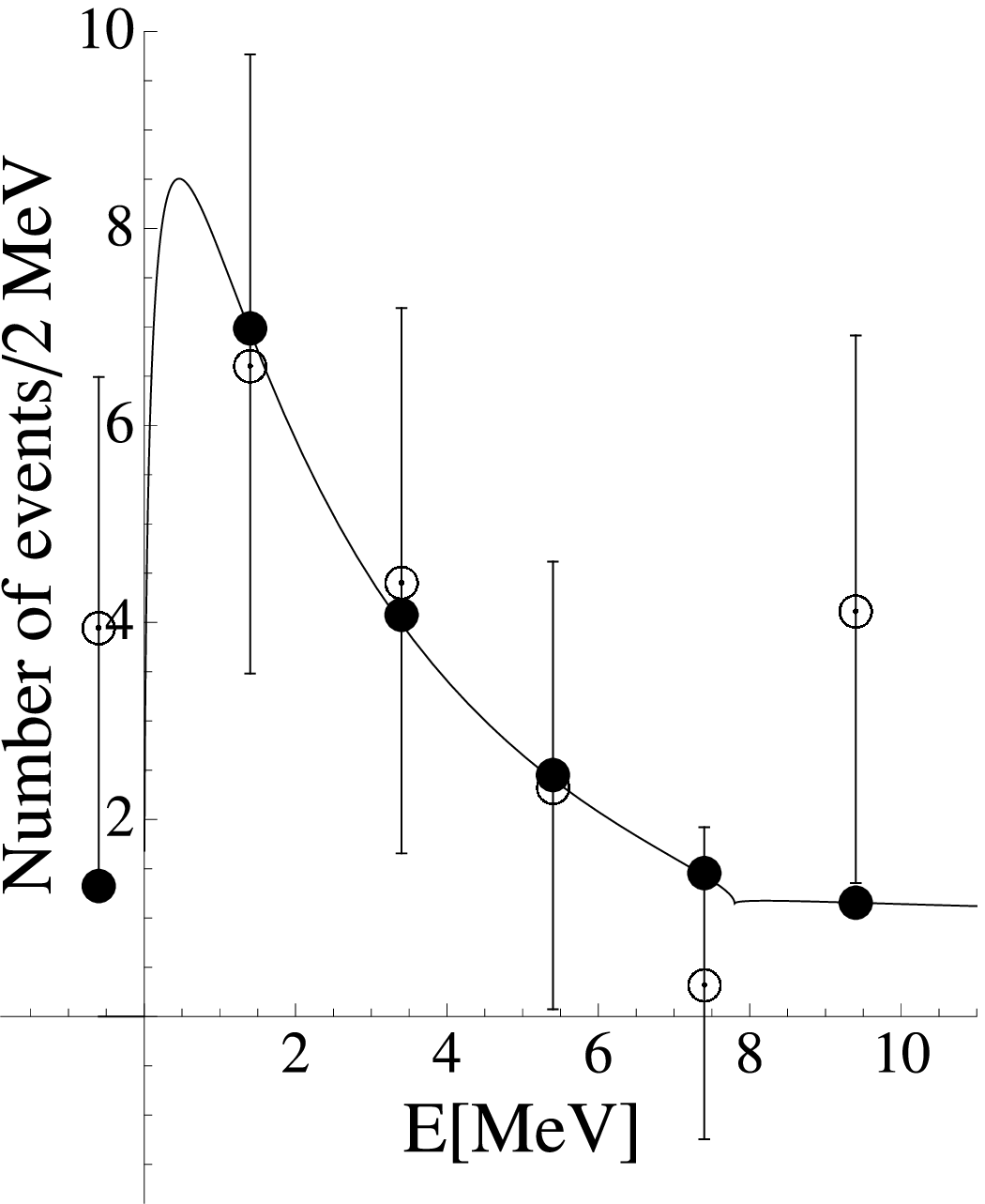,width=6cm}\\
\end{tabular}
\end{center}
\caption{The same as in Fig.~\ref{virboundfig} but for set~2 (upper plots)
and set~3 (lower plots).}\label{bellefig}
\end{figure}

The Belle data on the $D^0 \bar D^0 \pi^0$ mode \cite{Belle3875v2} can
be equally well described by both the virtual state and the bound state
in the $D\bar{D}^*$ system (set~1 and sets~2, 3 in
Table~\ref{bellesets} and plots in Figs.~\ref{virboundfig} and \ref{bellefig},
respectively). For set~1, the width $\varGamma_0$
mimics the $\gamma\psi'$ decay channel and is
fixed through the condition that $Br(\gamma\psi')\simeq
Br(\pi \pi J/\psi)$. However, the description of the Belle data on the $\pi^+
\pi^-J/\psi$ mode is remarkably poor for this set --- see
Fig.~\ref{virboundfig}. 
Besides that, the radiative width of about $1$ MeV seems to be suspiciously
large and, in any case, is not compatible with $\chi'_{c1}$ assumption.

We find therefore that a decent description of the Belle data on the $\pi^+
\pi^-J/\psi$ mode is only possible with a bound state (see sets~2 and 3 in
Table~\ref{bellesets} and in Fig.~\ref{bellefig}). 
Furthermore, because of a considerable contribution of the finite width
$\varGamma_0$ to the spectral density $w(E)$, its integral over the
near-threshold
region appears to be rather large (see Table~\ref{bellesets}) which indicates a
significant
admixture of the genuine charmonium state in the $X$ wave function.

\begin{figure}[t]
\begin{center}
\begin{tabular}{ccc}
\epsfig{file=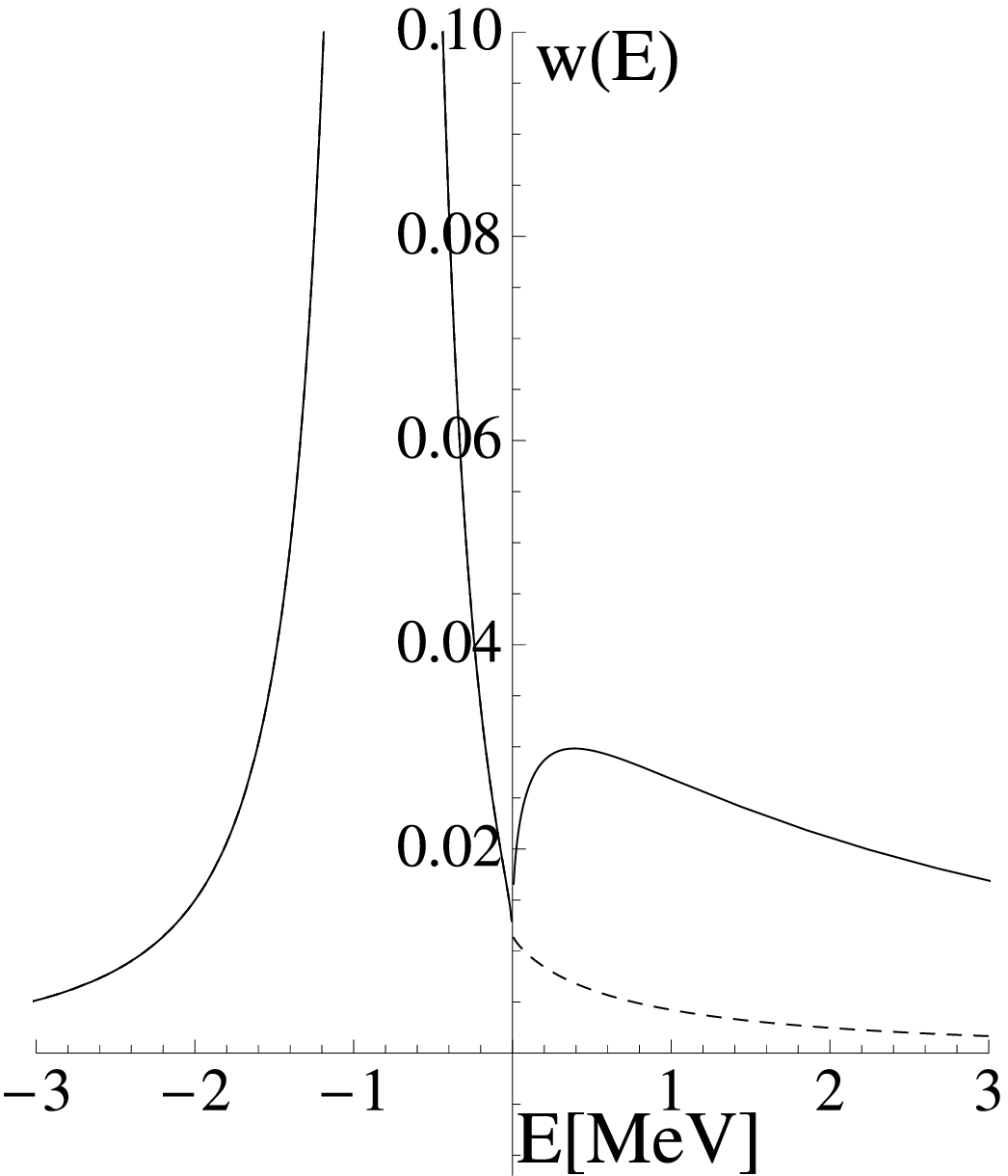,width=6cm}&\hspace*{2cm}&
\epsfig{file=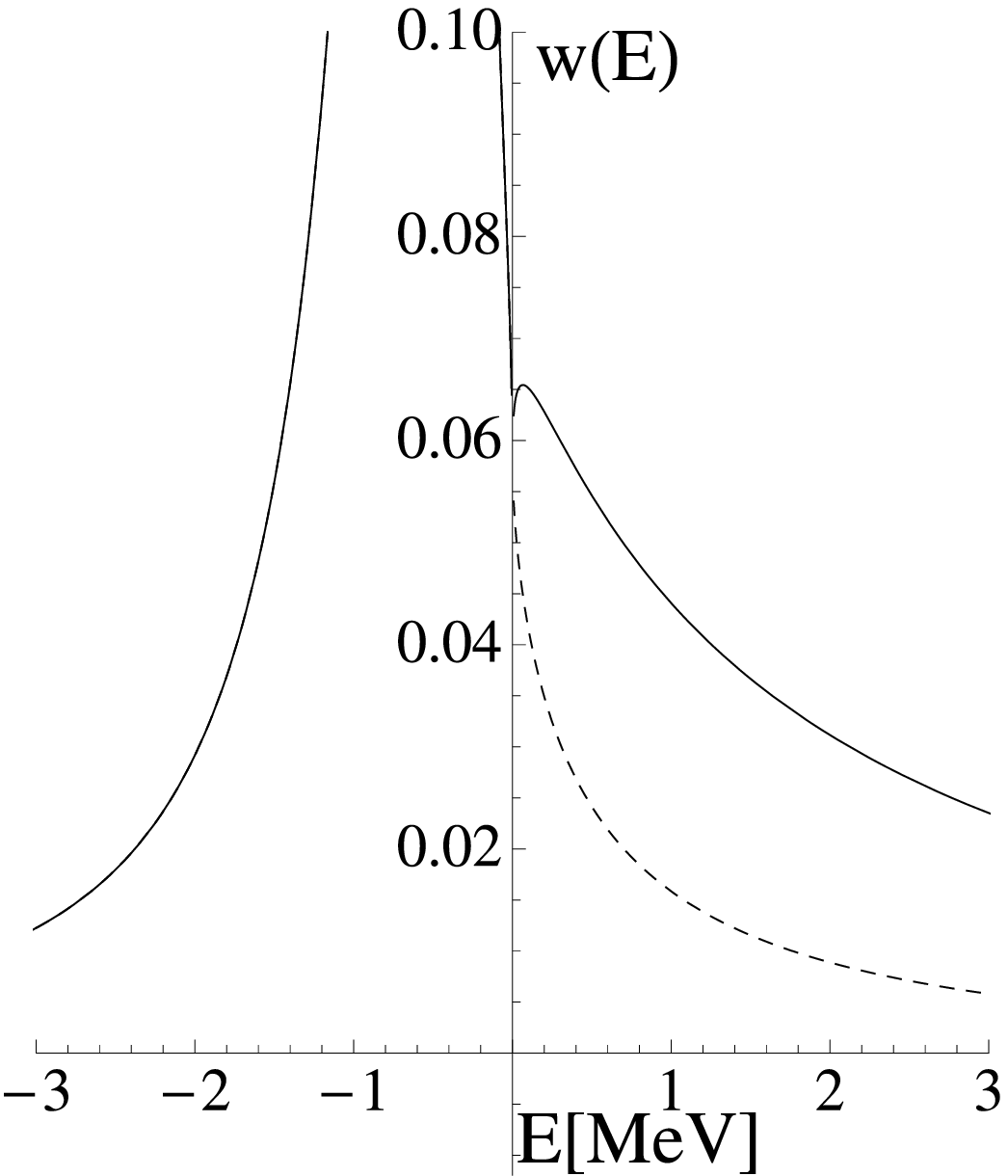,width=6cm}\\
\end{tabular}
\end{center}
\caption{Spectral density for set~2 (first plot) and set~3 (second plot).
The full spectral density $w(E)$ is plotted with the solid line, the
contribution $w_{inel}(E)$ of the non--$D \bar D^*$ channels to the spectral
density is shown with the dashed line. The functions $w(E)$ and 
$w_{inel}(E)$ coincide below threshold.}
\label{wfig}
\end{figure}

It is instructive to study the behaviour of the spectral density for
the bound--state case in more detail. It is plotted in Fig.~\ref{wfig}, where
the
contribution of non-$D \bar D^*$ modes is also shown which peaks at the
position of the bound-state mass. Using the relation between the \F~parameters
and the
effective range parameters established in Ref.~\cite{evi}, it is straightforward
to demonstrate that, in the limit of vanishing inelasticity, the spectral
density below the $D^0 \bar D^{*0}$  threshold becomes, similarly to
the inelastic
rate (\ref{inelrate})
proportional to a $\delta$-function,
\be
w(E) \to Z\delta(E-E_{\rm bound}),\quad E<0,
\ee
with the coefficient $Z$ being nothing but the famous $Z$-factor which
was introduced by
Weinberg in Ref.~\cite{weinberg} and which defines the probability to find a
bare state in the wave function of a physical bound state with the binding
energy $E_{\rm bound}$. So it is reasonable to define an integral 
over the near-threshold region:
\be
{\cal Z}=\int_{E_{\rm min}}^{E_{\rm max}} w_{inel}(E)dE,
\ee
with
\be
w_{inel}(E)=\frac{1}{2\pi|D(E)|^2}\varGamma(E).
\ee
Then the factor ${\cal Z}$ can be viewed as the $Z$-factor of our bound states
smeared due to the presence of the inelasticity and it takes the values:
\be
{\cal Z}=0.31~({\rm set~2}),\quad {\cal Z}=0.37~({\rm set~3}).
\ee

The values of the branchings $Br(B \to K\chi'_{c1})={\cal B}$ and $Br(B
\to K X)={\cal B}W$, as given in Table~\ref{bellesets}, agree with the
constrains imposed on them by experimental data and quoted in the beginning of
this section. The radiative decay width appears to be in a reasonable agreement
with quark model estimates (\ref{raddec}):
\be
\varGamma(\gamma\psi')=60~\mbox{keV (set 2)},\quad
\varGamma(\gamma\psi')=110~\mbox{keV (set~3)}.
\ee

Therefore, one can make the conclusion that the new Belle data favour the
$X(3872)$ to be a mixture of a genuine charmonium and a dynamically generated
molecule-like state which appears to be a bound state of the $D^0 \bar D^{*0}$
system. 

\subsection{BaBar Collaboration data}

In the data analysis procedure, similarly to the Belle data case, the formulae
for the number-of-events distributions were taken to be:
\be
N^{\pi \pi J/\psi}_{\rm BaBar}(E)=5 \,  {\rm [MeV]} \left(\frac{93.4}{8.4 \cdot
10^{-6}}\right)\frac{dBr(B \to K \pi^+ \pi^- J/\psi)}{dE},
\label{rhonoebabar}
\ee
for the $\pi^+\pi^- J/\psi$ mode (bin size is $5$ MeV, number of events is
$93.4$, and $Br(B \to K \pi^+\pi^- J/\psi)=8.4 \cdot 10^{-6}$ --- see
Ref.~\cite{BaBar2}), and
\be
N^{D^0 \bar D^{*0}}_{\rm BaBar}(E)=2.0 {\rm [MeV]} \left(\frac{33.1}{1.67 \cdot
10^{-4}}\right)\frac{dBr(B \to K D^0 \bar D^{*0})}{dE},
\label{dnoebabar}
\ee
for the $D^0 \bar D^{*0}$ mode (bin size is $2$ MeV, number of events is
$33.1$, and $Br(B \to K D^0 \bar D^{*0})=1.67 \cdot 10^{-4}$; all $D^0 \bar
D^{*0}$ modes are included --- see Ref.~\cite{Babarddpi}). The
signal--background
interference is taken into account in the same manner as for the Belle data --- see Eq.~(\ref{tot2}), with the factor $0.62$ omitted.  

The resolution function for the $\pi^+\pi^-
J/\psi$ channel is taken in the form of
a Gaussian with the fixed resolution scale being $4.38$ MeV \cite{BaBar2}. As
to the $D \bar D^*$ resolution, it is described by the BaBar Collaboration as
a very complicated function, and it is not available in public domain. In
the present analysis we take, with corresponding reservations, this resolution
also to be Gaussian, with the resolution scale of $1$ MeV.

\begin{figure}
\begin{center}
\begin{tabular}{ccc}
\epsfig{file=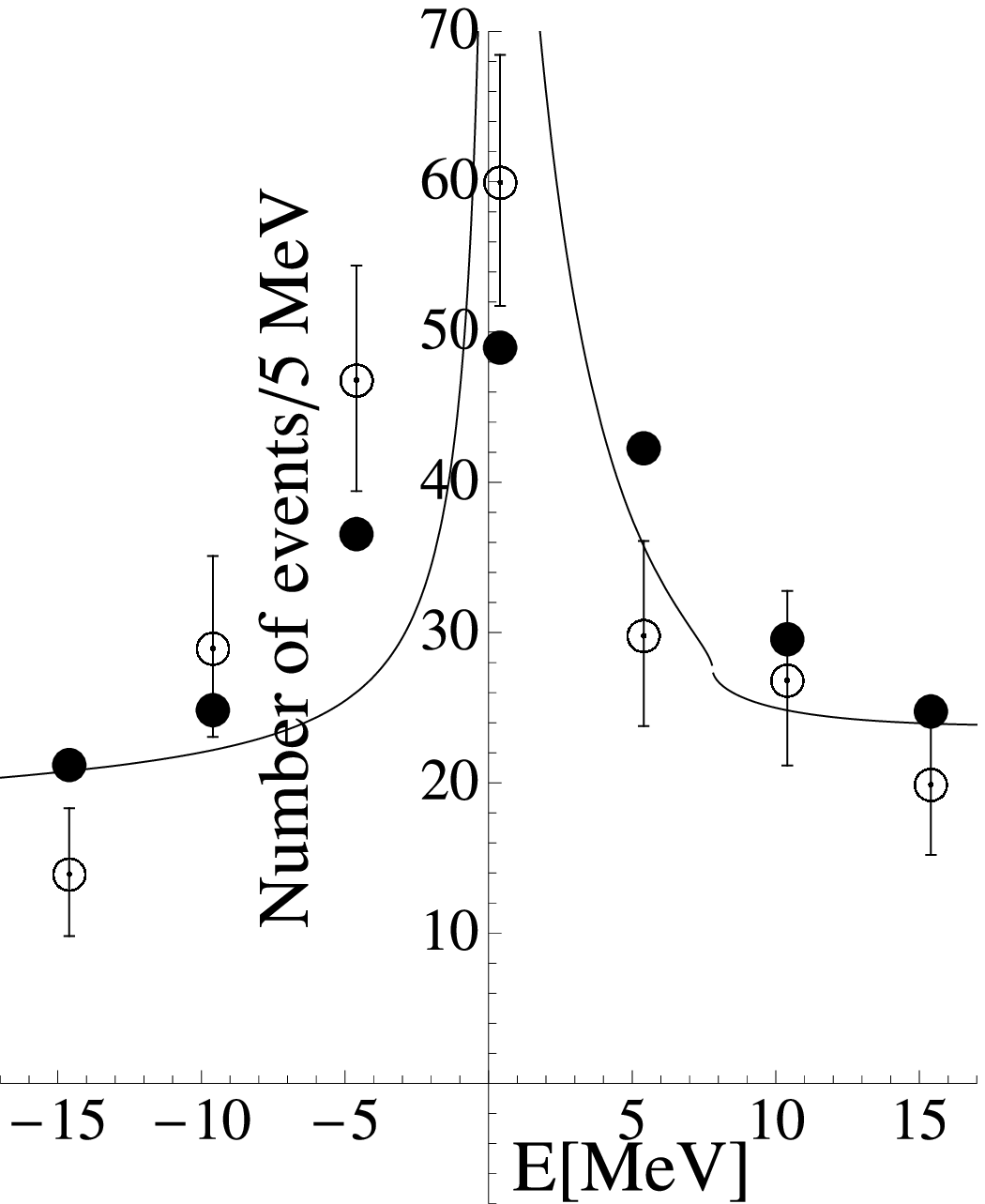,width=6cm}&\hspace*{1cm}&
\epsfig{file=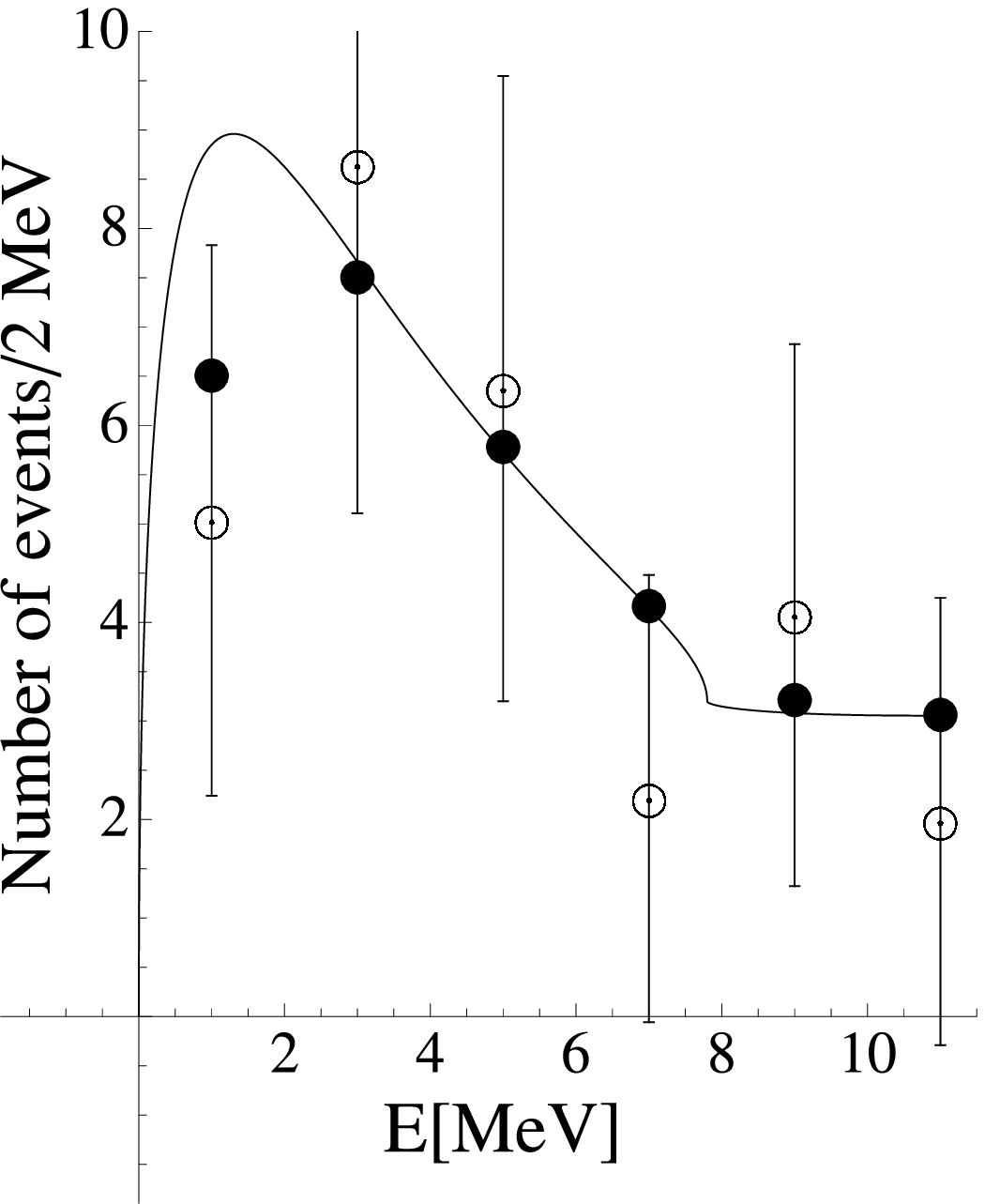,width=6cm}\\
\epsfig{file=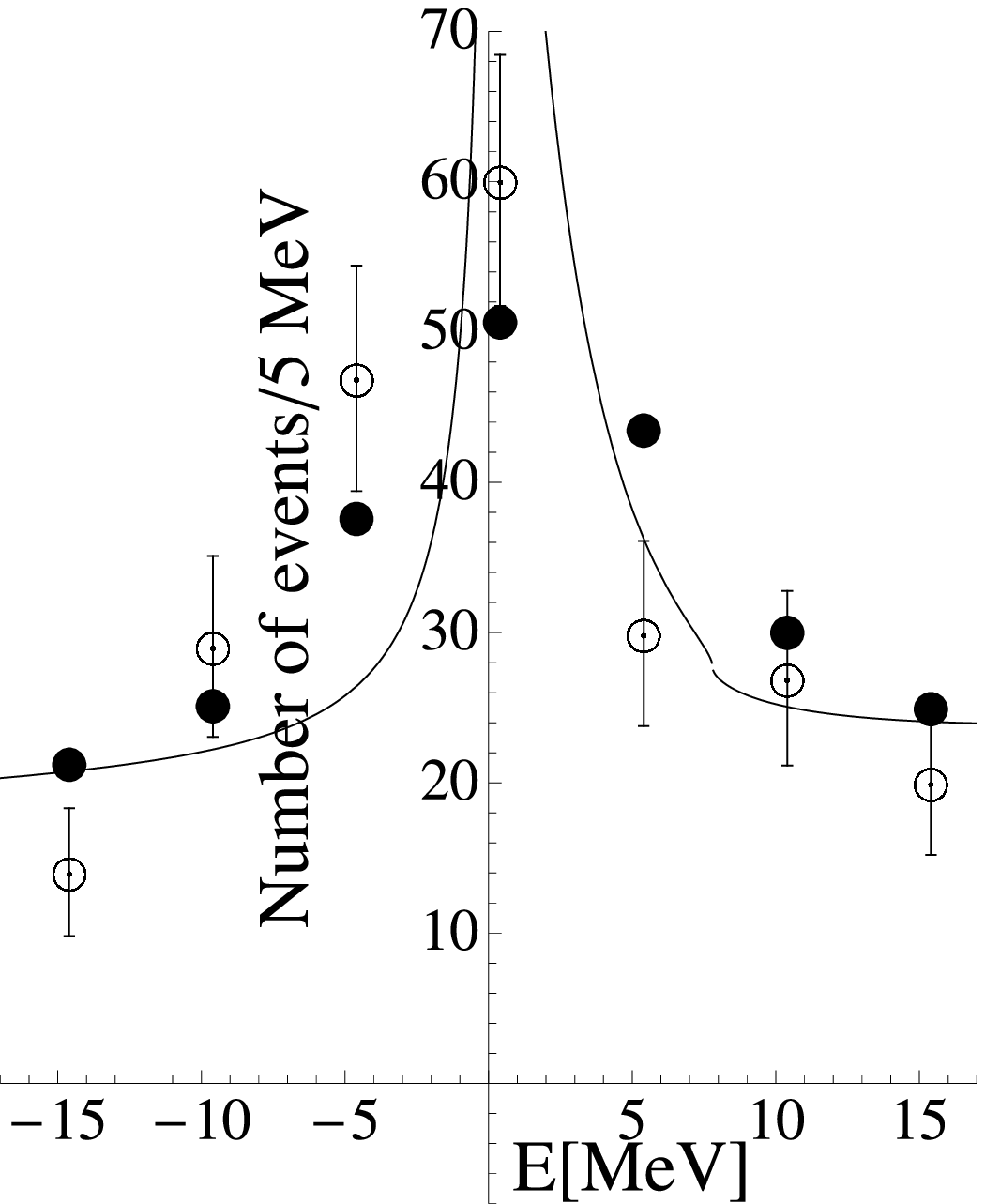,width=6cm}&&
\epsfig{file=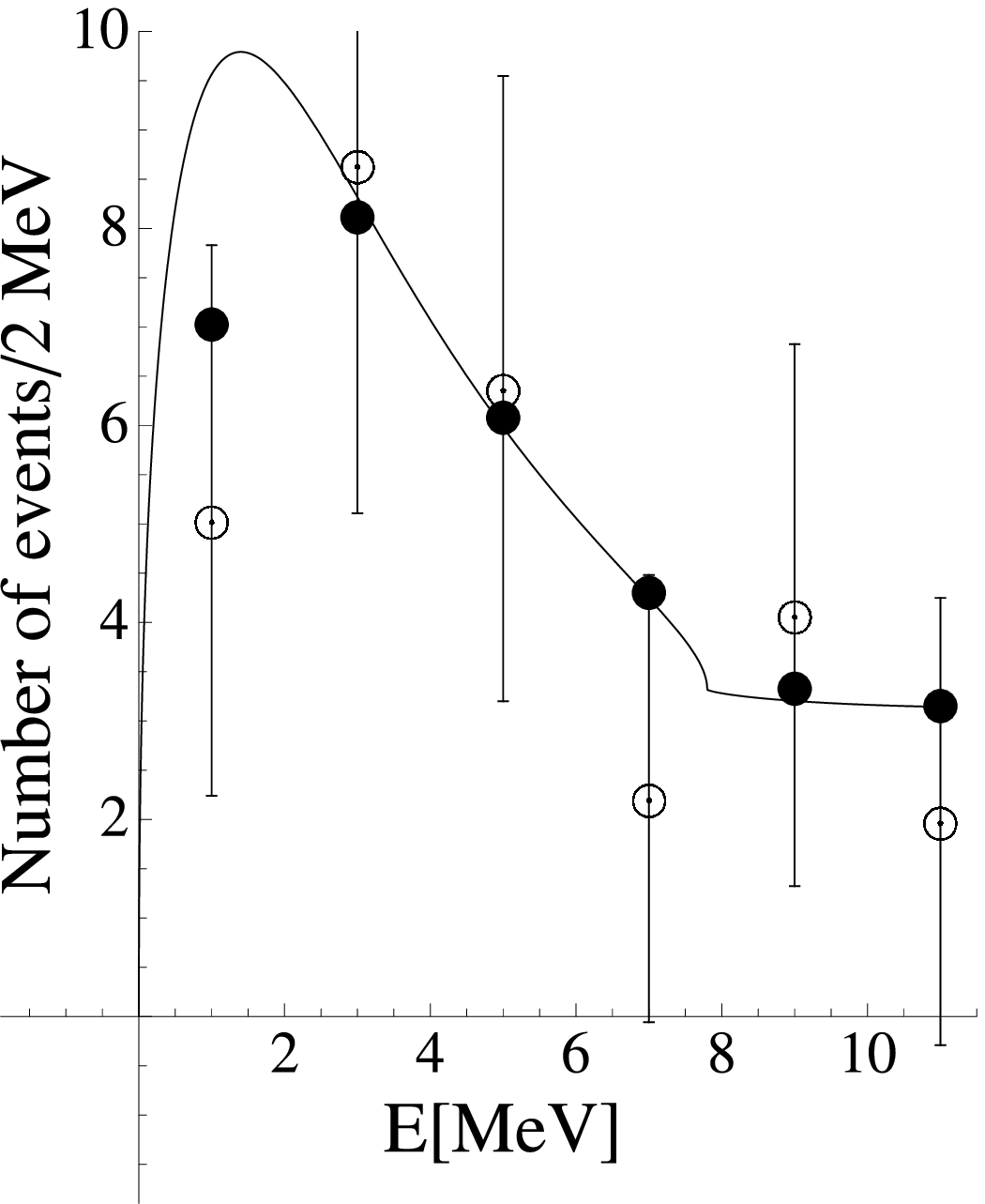,width=6cm}\\
\end{tabular}
\end{center}
\caption{Differential rates for the $\pi^+ \pi^-J/\psi$ channel (left plots) and
$D^0 \bar{D}^{*0}$ channel (right plots) (see Eqs.~(\ref{rhonoebabar}) and
(\ref{dnoebabar}), respectively) with the parameters given by set~4 (upper
plots) and set~5 (lower plots). Parameters from these sets are presented in
Table~\ref{babarsets}. The distributions integrated over the bins, with 
resolution function taken into account, are shown
as filled dots, experimental data (see Refs.~\cite{BaBar2,Babarddpi}) are
given as open dots with error bars.}\label{babfig1}
\end{figure}

\begin{figure}
\begin{center}
\begin{tabular}{ccc}
\epsfig{file=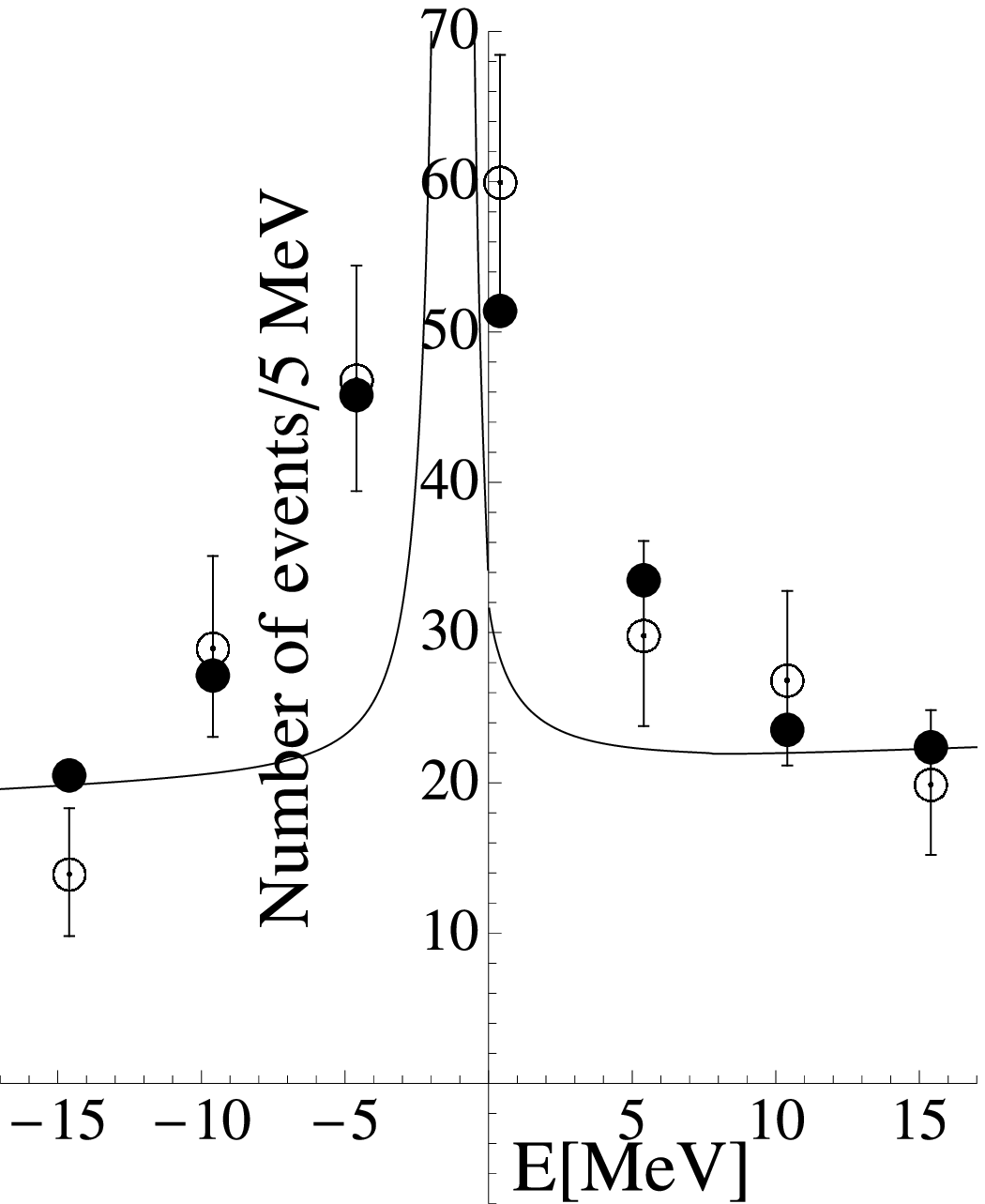,width=6cm}&\hspace*{1cm}&
\epsfig{file=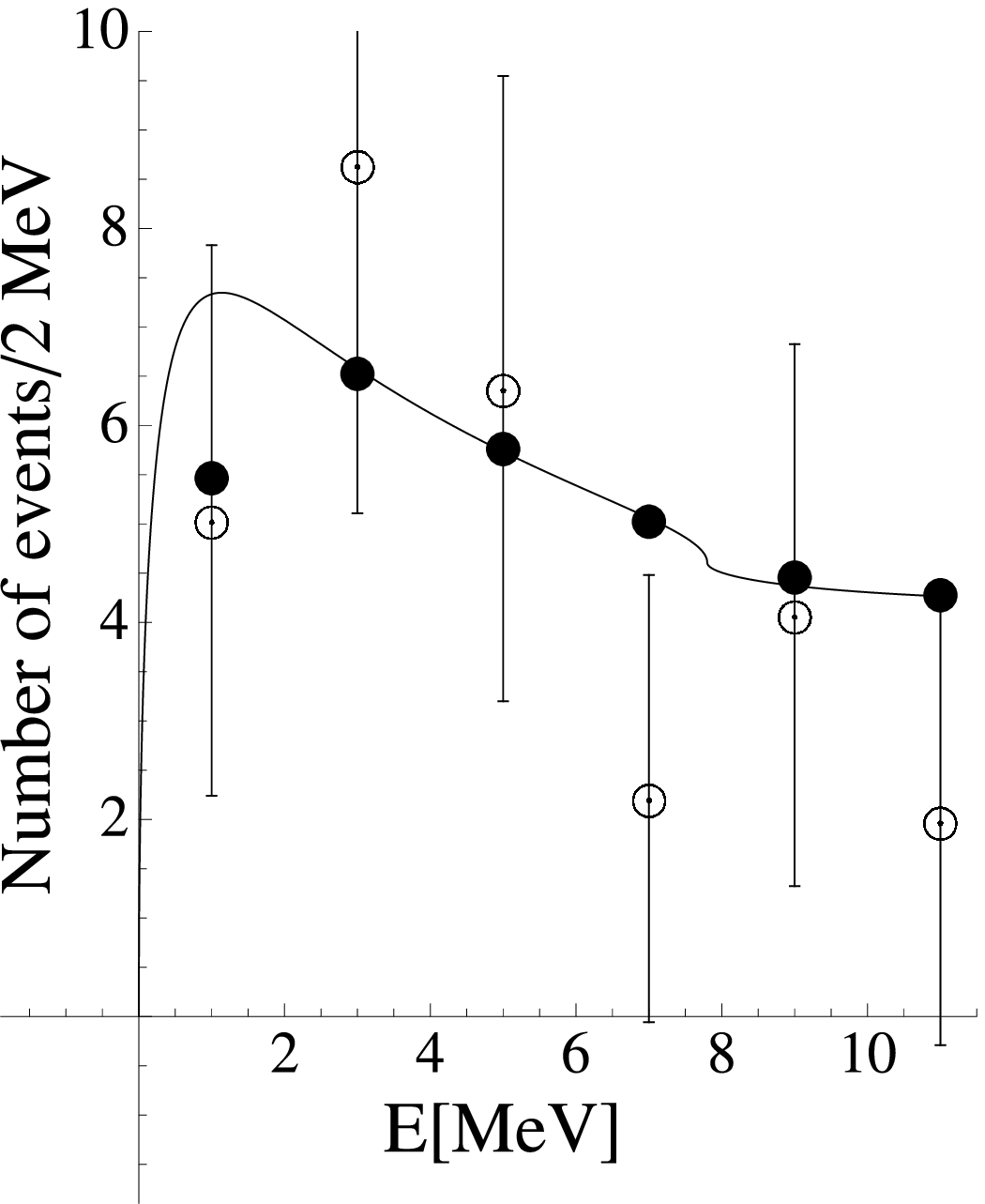,width=6cm}\\
\epsfig{file=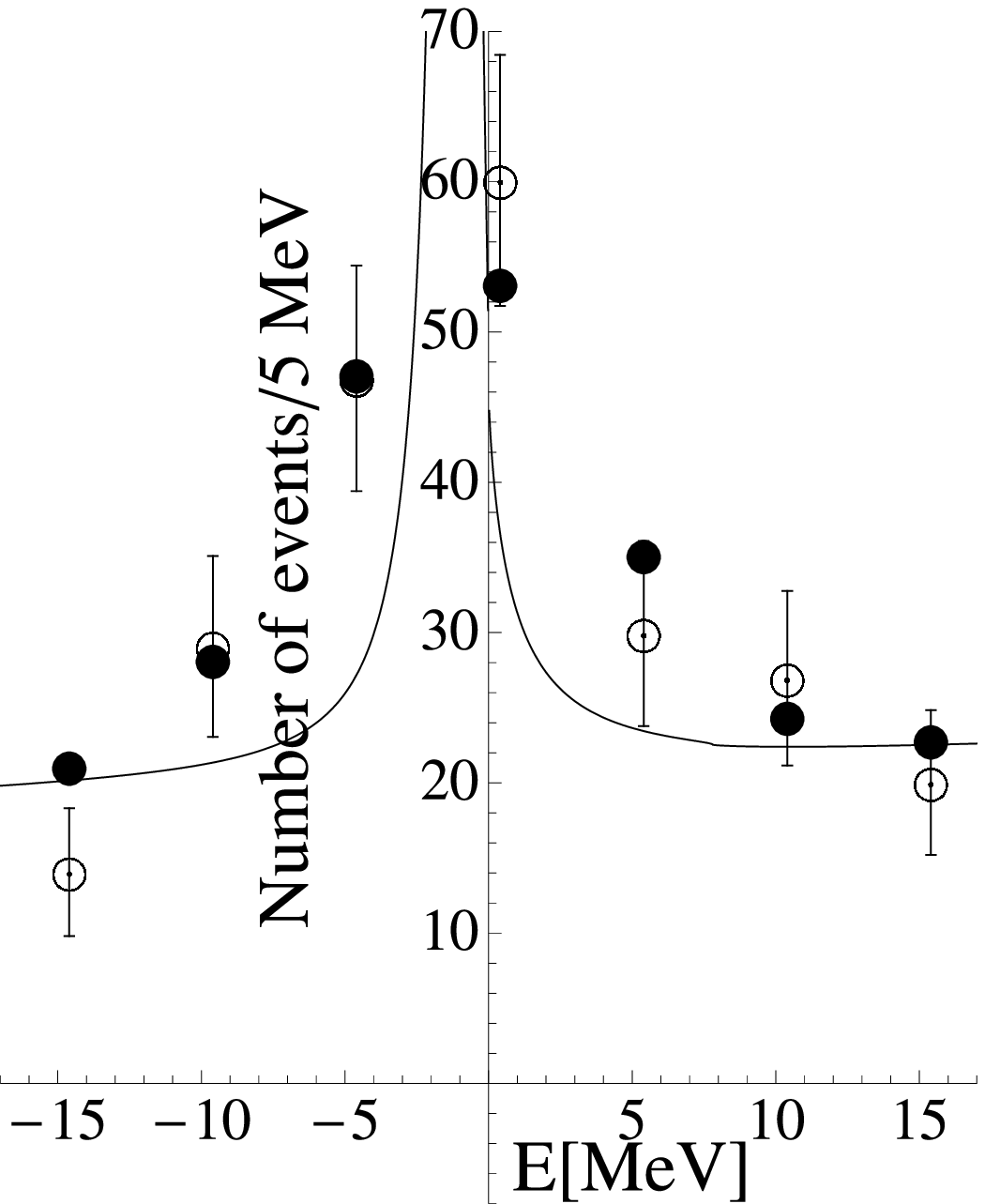,width=6cm}&&
\epsfig{file=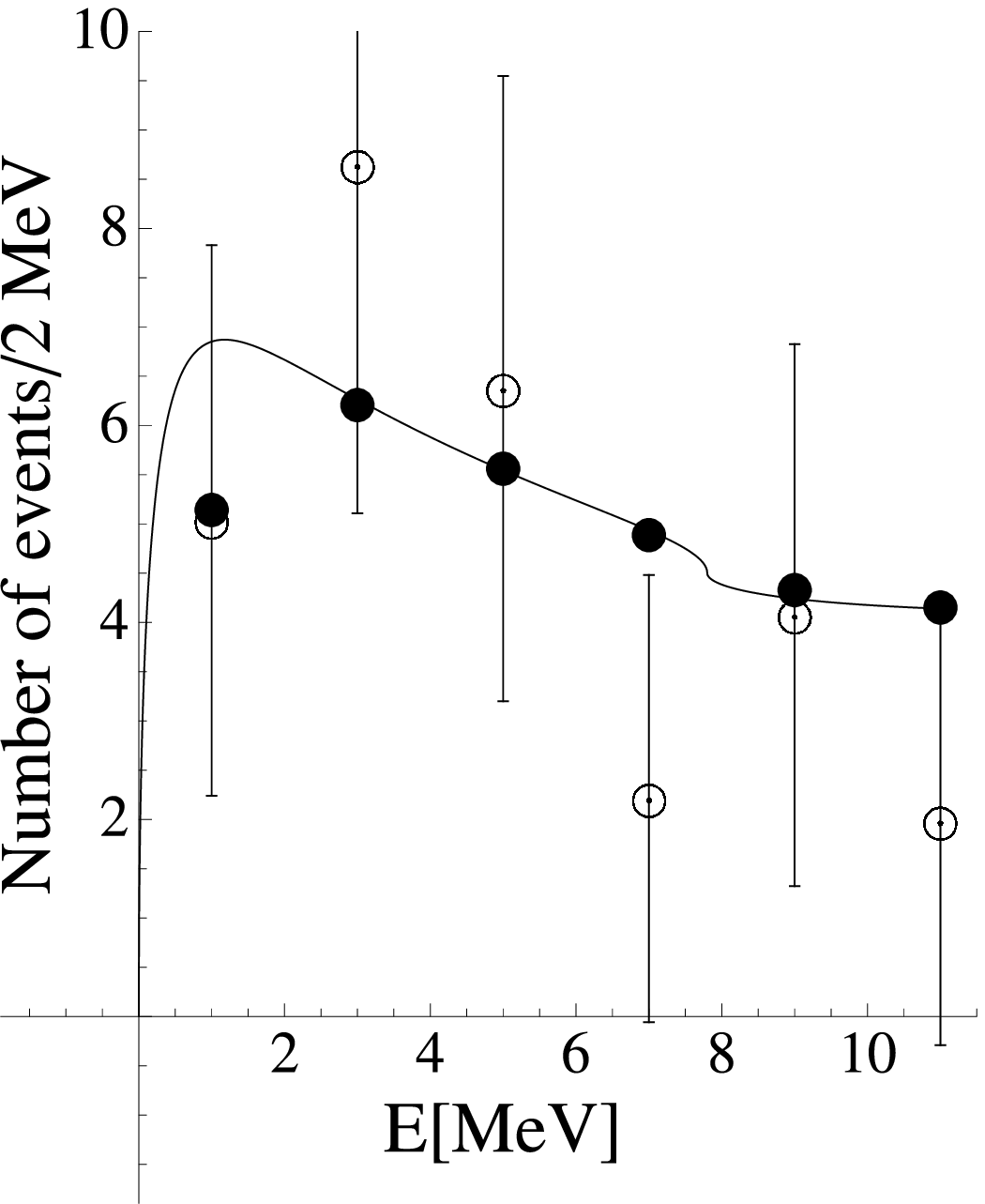,width=6cm}\\
\end{tabular}
\end{center}
\caption{The same as in Fig.~\ref{babfig1} but for set~6 (upper plots)
and set~7 (lower plots).}\label{babfig2}
\end{figure}

The BaBar $D^0 \bar D^{*0}$ data \cite{Babarddpi} are very similar to the old
Belle ones \cite{Belle3875}, while the
$\pi^+\pi^- J/\psi$ peak in Ref.~\cite{BaBar2} is moved a bit to the left in
comparison with the old BaBar data on the same reaction, and the peak width has
decreased around 25\% due to a better resolution. One expects therefore, that
the $D^0 \bar D^{*0}$ data are better described as a
virtual state, while the $\pi^+\pi^- J/\psi$ data complies better with the bound
state. 
Correspondingly, we employ two different analysis strategies. First, we 
reconcile the $\pi^+\pi^-J/\psi$ and $D^0 \bar D^{*0}$ peaks with each other, as
it was done in \cite{recon,YuSKconf} (sets~4 and 5).
The second strategy is to find the best overall description of the both data
sets (sets~6 and~7). The parameters for these sets are
given in Table~\ref{babarsets}, and the differential rates are shown in
Figs.~\ref{babfig1} and~\ref{babfig2}.

\begin{table}[t]
\caption{The sets of the \protect{\F} parameters for the
BaBar data from Refs.~\cite{BaBar2,Babarddpi}.}
\begin{ruledtabular}
\begin{tabular}{ccccccccccc}
Set&$\varGamma_0$&$g$&$E_f$, MeV&$f_\rho$&$f_\omega$&
${\cal B}\cdot 10^4$&$\phi$&$W$&${\cal B}W\cdot 10^4$&$a$, fm\\
\hline
4&1.0&0.225&-9.7&0.0065&0.0360&3.9&$113^0$&0.24&1.8&$-4.9-i1.6$\\
5&2.0&0.145&-6.0&0.0040&0.0230&3.6&$109^0$&0.34&0.8&$-3.9-i2$\\
6&1.0&0.080&-8.4&0.0002&0.0010&5.7&$0^0$&0.58&3.3&$2.2-i0.3$\\
7&2.0&0.090&-9.0&0.0005&0.0029&5.5&$0^0$&0.53&2.9&$3.3-i0.7$\\
\end{tabular}
\end{ruledtabular}
\label{babarsets}
\end{table}

The reconciling procedure yields a virtual state, similar to the one found in
Refs.~\cite{recon,YuSKconf}. The admixture of the charmonium is not large for
these
sets and, again, the radiative decay width $\varGamma(\gamma \psi')$
seems to be too large for the charmonium assignment:
\be
\varGamma(\gamma \psi')=800~\mbox{keV (set~4)},\quad
\varGamma(\gamma \psi')=500~\mbox{keV (set~5)}.
\ee
The overall description of the data seems to be not too bad, as the present
resolution cannot confirm or rule out the $\pi^+\pi^- J/\psi$ cusp
scenario.

Solutions $6$ and $7$, obtained from the overall fit to the BaBar data, clearly
prefer the bound--state, similarly to the ones given by sets~2 and~3 for the
Belle data. The
charmonium admixture is even larger than for the Belle version (see
Table~\ref{babarsets}), and the
$\varGamma(\gamma \psi')$ width,
\be
\varGamma(\gamma\psi')=25~\mbox{keV (set~6)},\quad
\varGamma(\gamma\psi')=60~\mbox{keV (set~7)},
\ee
is a bit small as compared to the estimates (\ref{raddec}). The $\pi^+\pi^-
J/\psi$ data are described better than in the virtual--state version,
while the description of the $D^0 \bar D^{*0}$ data is rather poor (it remains
an open question either it is genuine, or is due to our ill-starred guess on
the BaBar $D^0 \bar D^{*0}$ resolution function). 

Therefore, we find that the BaBar data are more compatible with the assumption
of the $X(3872)$ being a virtual state of a dynamical nature. As to
the charmonium admixture, we can only state that it is small.
Indeed, as shown in Ref.~\cite{evi}, in the
case of small values of $W$, the model independent \F~analysis does
not allow one to draw any conclusions on the binding mechanism, that is to
distinguish between
$t$-channel meson exchange forces or short-ranged $s$-channel forces
due to coupling of bare states to the hadronic channel. One can only
state that the properties of the resonance are given mainly by the
hadronic continuum contribution, and the state is mostly of
a dynamical (molecular) nature.

We conclude this section with the comment on the paper~\cite{ch2}, where
the \F~fits were performed similar to ours, and the pole structure of the
\F~amplitude was studied. 
Conclusions on the nature of the $X$ were drawn
in Ref.~\cite{ch2} based on the pole counting procedure developed in
Ref.~\cite{morgan}: two near-threshold poles correspond to a large admixture of
a bare state in the resonance wave function, while a single near-threshold pole
indicates a dynamical nature of the resonance.
Strictly speaking, the Riemann surface for the $X$ case is much more complicated
than the one assumed in Ref.~\cite{ch2}, due to the presence of many--body cuts
(caused by the $\pi^+\pi^- J/\psi$ and $\pi^+\pi^-\pi^0 J/\psi$ modes). However,
the pole--counting procedure should yield, qualitatively, the same result as a
more rigorous method based on the spectral density calculations employed here,
which allows, {\it inter alia}, to estimate quantitatively the bare state
admixture (for more details on the interrelation between the pole-counting and
spectral density behaviour see Ref.~\cite{evi}). For the fits with reasonably
small values of the factor ${\cal B}$ two near-threshold poles were found
in Ref.~\cite{ch2}, signalling a large admixture of the genuine charmonium,
similarly to our Belle parameter sets. Note, however, that 
the fits presented in Ref.~\cite{ch2} are the overall ones: the Belle and BaBar
data were fitted simultaneously. As a result, rather poor description of
the BaBar $D^0 \bar D^{*0}$ data was obtained, reflecting incompatibility of
the new Belle and BaBar data.

\section{Comment on the $D^*$ finite width}\label{finwidth}

The possibility of the bound-state solution brings on board one more important
question. Namely, in the present analysis, the $D^{*0}$-meson was assumed to be
stable. As argued in Ref.~\cite{recon}, account for a small finite width of the
$D^{*0}$ does not
change the $D^0 \bar D^{*0}$ line-shape in the case of the virtual state while,
for a bound state, the effects of the finite width could be pronounced, as shown
in Ref.~\cite{braatenfw}.
A refined treatment of the finite width is in progress now \cite{ournew},
while here we estimate these effects using a simple ansatz
suggested in Ref.~\cite{woolly} and re-invented in Ref.~\cite{braatenfw}. The
recipe is to
make the following replacement in the expressions for the $D^0 \bar D^{*0}$
momentum
entering the formulae for differential rates:
\be
\Theta (E)k_1(E) \to \sqrt{\mu_1}\sqrt{\sqrt{E^2+\varGamma_*^2/4}+E},
\label{keff}
\ee
and
\be
\Theta (-E)\kappa_1(E) \to
\sqrt{\mu_1}\sqrt{\sqrt{E^2+\varGamma_*^2/4}-E},
\label{kappaeff}
\ee
where $\varGamma_*$ is the width of the $D^{*0}$-meson. It can be shown
\cite{ournew} that these formulae are valid if the resonance is well-separated
from the three-body threshold (the $D^0 \bar D^{*0} \pi^0$ threshold in our
case), and
the zero-width limit is readily reproduced as $\varGamma_*\to 0$.

\begin{figure}
\begin{center}
\begin{tabular}{ccc}
\epsfig{file=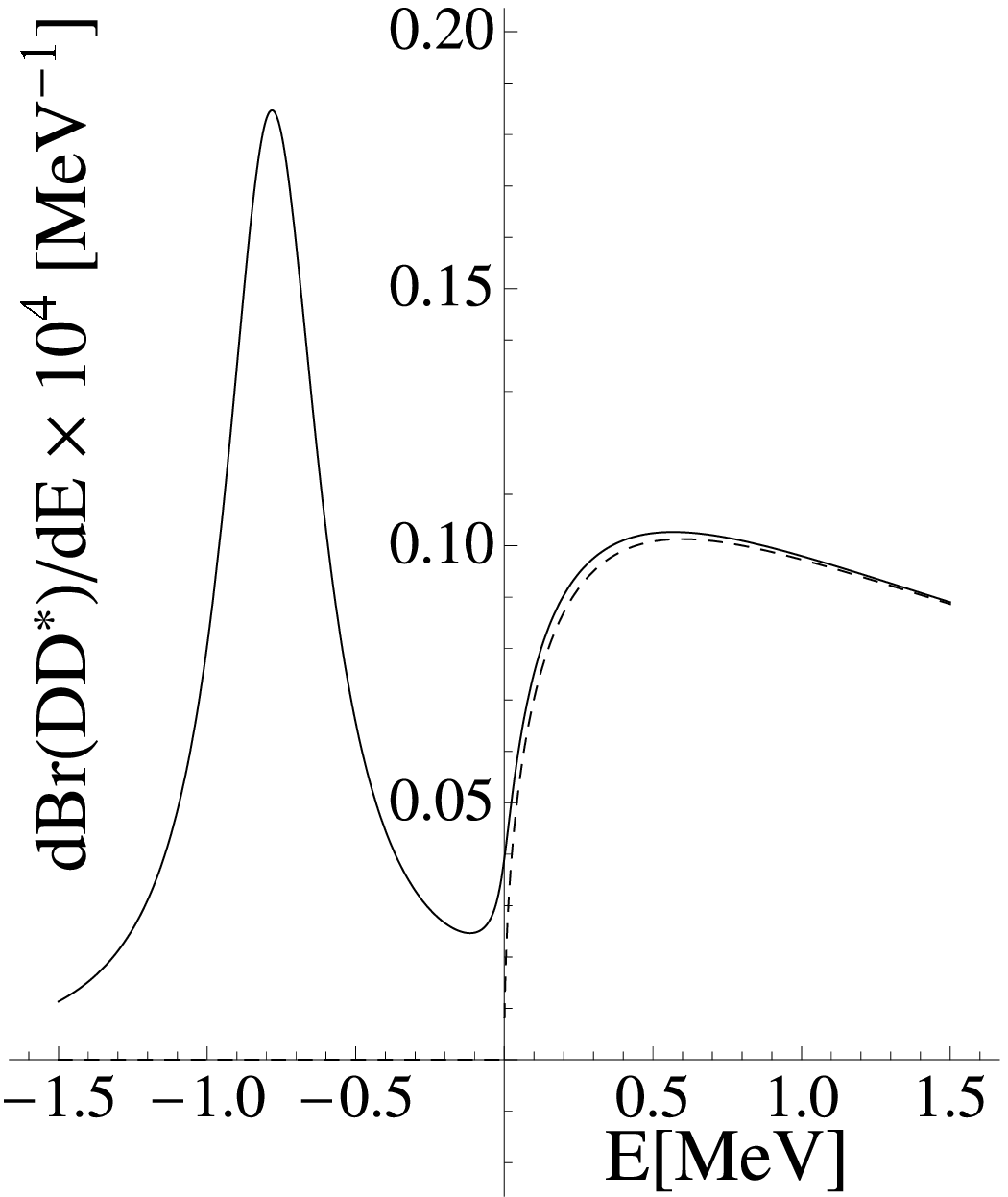,width=6cm}&\hspace*{2cm}&
\epsfig{file=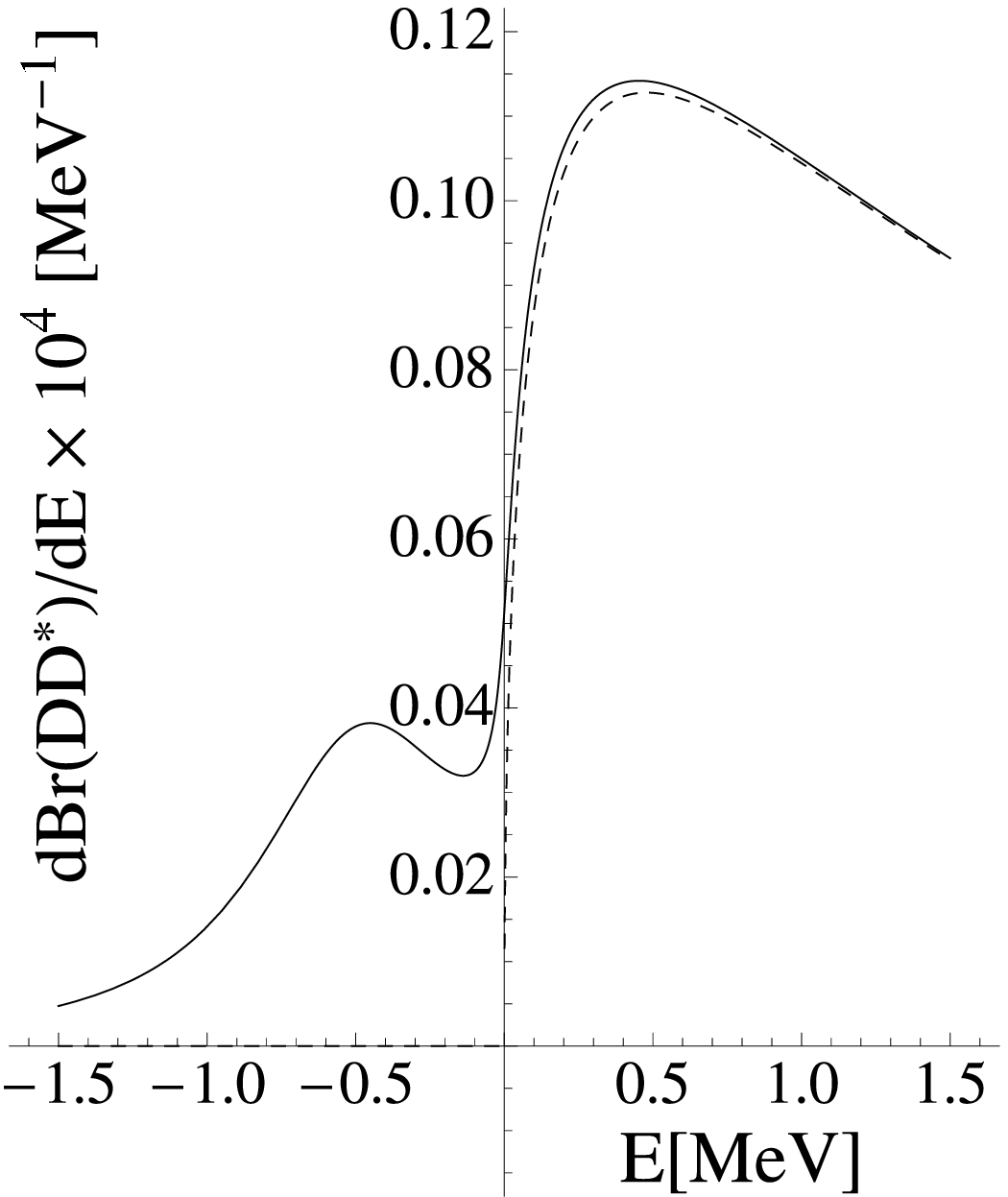,width=6cm}\\
\epsfig{file=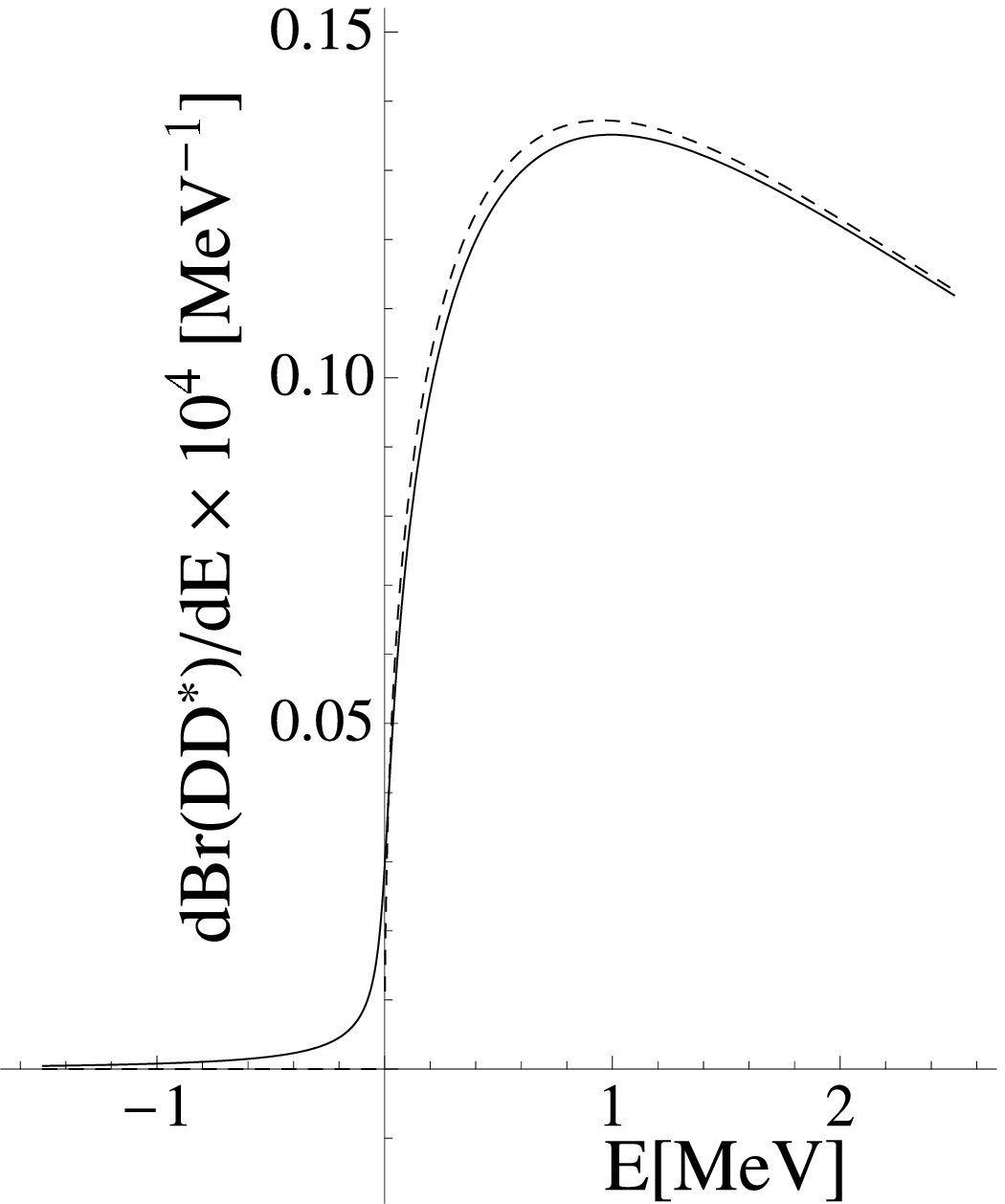,width=6cm}&&
\epsfig{file=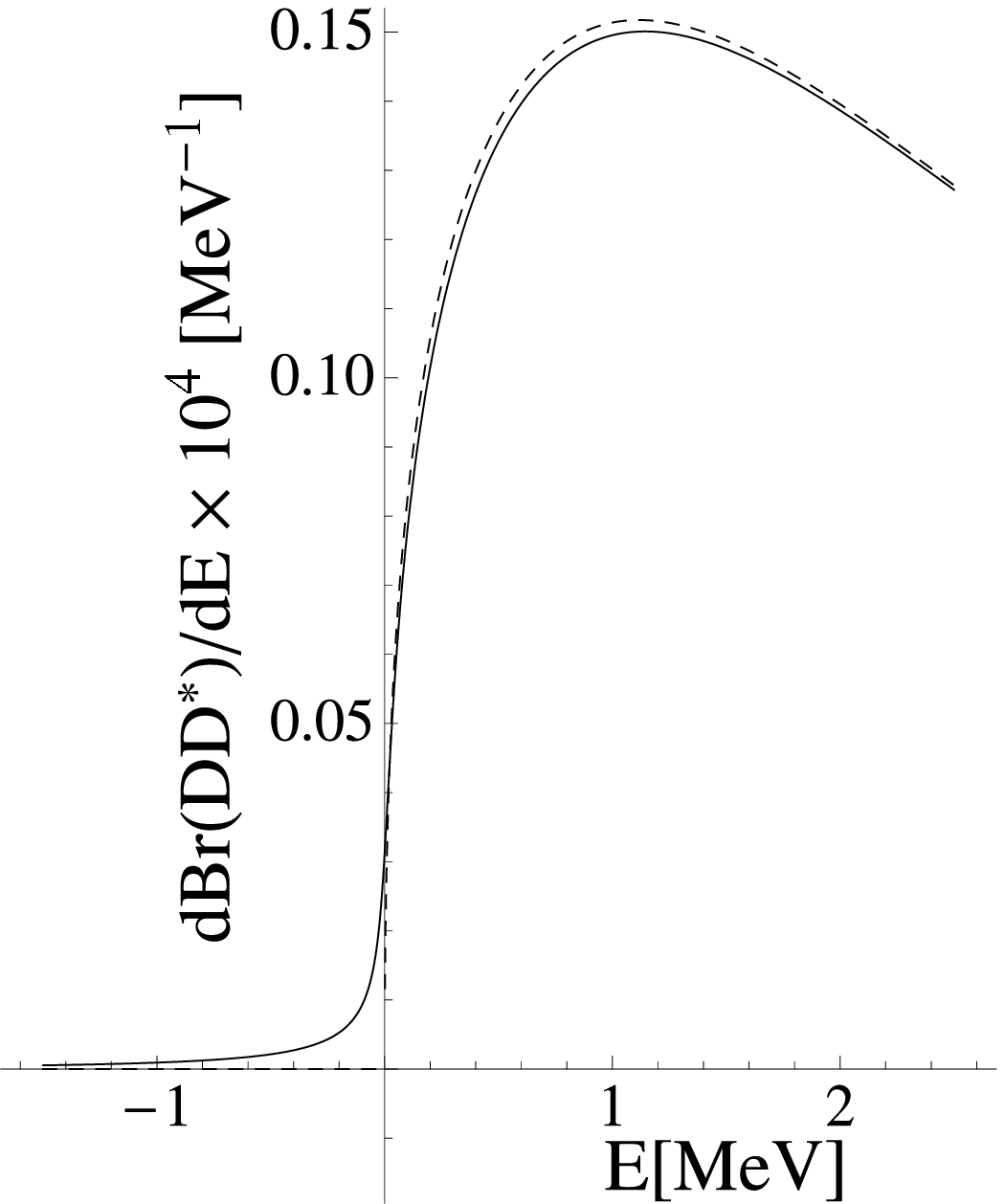,width=6cm}\\
\end{tabular}
\end{center}
\caption{Upper plots: the differential rates for the 
$D^0\bar{D}^{*0}$ channel
given by the
\protect{\F}~formula with the finite width $\varGamma_*$ included (solid lines)
and with the zero width (dashed lines) for parameters sets~2 (left) and 3
(right).
Lower plots: the same as in the upper plots but
for parameters sets~4 and 5.}\label{fw}
\end{figure}

To access the role of the finite $D^{*0}$ width we evaluate the $D^0 \bar
D^{*0}$ differential rates, with $\varGamma_*=63$ keV, for the sets
2 and 3 (bound state scenario for the Belle data) and for the sets 4 and 5
(virtual state scenario for the BaBar data) and plot them, together with the
zero-width rates, in Fig.~\ref{fw}. In particular, in this figure, we
show our theoretical curves given by expressions (\ref{DD}), with the
replacement (\ref{keff}), (\ref{kappaeff}), and with the \F~parameters from
the corresponding tables, without signal-background interference and not
smeared with the resolution functions.

As seen from Fig.~\ref{fw}, the virtual-state solutions are not affected by
the finite $D^{*0}$ width at all (lower plots). In the meantime, the bound-state excitation curves are not affected either, in the above-threshold region, and a
non-negligible $D^0 \bar D^0 \pi^0$ peak is developed around the bound-state
mass position (upper plots). 

The bound-state peak resides at about $-0.75$ MeV for set~2 and at about
$-0.5$ MeV for set~3, so the only effect expected is an increase of the
number of events in the first near-threshold bin. Indeed, for the Belle
bound-state solutions, we have calculated the ratio $\tilde{N}_i/N_i$ of
the number of events in the first ($i=1$) and second ($i=2$) non-empty Belle
bins, with ($\tilde{N}_i$) and without ($N_i$) inclusion of the finite width:
\begin{eqnarray*}
\tilde{N}_1/N_1=4.31,\quad \tilde{N}_2/N_2=1.01\quad\mbox{(set~2)},\\
\tilde{N}_1/N_1=1.99,\quad \tilde{N}_2/N_2=1.00\quad\mbox{(set~3)},
\end{eqnarray*}
where the ratios above are calculated without resolution and signal--background
interference. Clearly, a large value of the ratio $\tilde{N}_1/N_1$ 
does not cause problems, as Belle 
bound-state solutions underestimate the number of events in the lowest 
bin only (see Fig.~\ref{bellefig}), and the number of events in higher bins 
is not affected by the finite-width effect.

Thus a considerable number of the $D^0 \bar D^{*0}$ events is to appear below
the nominal $D^0 \bar D^{*0}$ threshold in the bound-state case. In the
meantime,
the present experimental situation does not allow one to identify the
bound-state peak. This must be attributed to the
peculiarities of the data analysis: both BaBar and Belle
Collaborations assume that the $D^0 \bar D^0 \pi^0$ events come from the $D^0
\bar D^{*0}$, distorting in such a way the kinematics of the below-threshold
events and feeding artificially the above-threshold region at the expense of the
below-threshold one.    

In a quite recent paper \cite{braatenfw2} the $D^0 \bar D^0
\pi^0$ distributions were obtained with the recipe
(\ref{keff}), (\ref{kappaeff}), and, in order to describe the $D^0 \bar D^{*0}$
data, the above-mentioned kinematical distortion was corrected with some
feedback from data processing arrangements. As a result, a nice description of the
$D^0 \bar D^{*0}$ Belle data was obtained with a bound-state solution (the
description of the BaBar data is rather poor in Ref.~\cite{braatenfw2}, quite
similar to our BaBar bound--state solutions~6 and 7). Notice, however, that the
solutions of Ref.~\cite{braatenfw2} differ from
ours in several respects. First, the data on $\pi^+\pi^- J/\psi$ and $D^0
\bar D^{*0}$ modes are analysed separately in Ref.~\cite{braatenfw2}, so that it
remains unclear whether the presented solutions 
provide a tolerable overall fit. Second, there is no extra width $\varGamma_0$
in the fits describing the $D^0 \bar D^{*0}$ mode (in fact, there is no
inelasticity at all in the best fits for the $D^0 \bar D^{*0}$ data). Besides that, the scattering length approximation for the $D^0 \bar
D^{*0}$ scattering amplitude is employed in Ref.~\cite{braatenfw2}, which is not
adequate for our solutions.

Finally, related to the question of the finite $D^{*0}$ width is the
problem of the
interference in the decay chains
$X\to D^0 \bar D^{*0} \to D^0 \bar D^0 \pi^0$ and $X\to \bar D^0 D^{*0}  \to D^0
\bar D^0
\pi^0$. According to the estimates made in Ref.~\cite{voloshinint}, the
interference effects could enhance
the below-threshold $D^0 \bar D^0 \pi^0$ rate up to two times, however the
effect is much more moderate above threshold \cite{ournew}. As to the  $X\to D^0
\bar D^{*0} \to D^0 \bar D^0 \gamma$ and $X\to \bar D^0 D^{*0}  \to D^0 \bar D^0
\gamma$ decay chains, these are shown to interfere
destructively \cite{voloshinint}.
The proper
account for the interference cannot be done in the over-simplified framework 
presented here, as this effect is
to be included in the coupled-channel scheme from the very beginning
\cite{ournew}.

\section{Discussion and Outlook}\label{discsect}

The present analysis was prompted mainly by two recent experimental results:
the discovery of the new $\varGamma(\gamma \psi')$ mode of the $X(3872)$
\cite{BaBarrad} and the new Belle data \cite{Belle3875v2} on the $D^0 \bar
D^{*0}$
mode. We have arrived at conclusions different from the ones of the papers
\cite{recon,YuSKconf}.
It is instructive to discuss in detail the relation between these new results
and the previous ones.

To begin with, the $\varGamma(\gamma \psi')$ mode is to be included in the
analysis. As the corresponding rate is comparable to the $\pi^+\pi^- J/\psi$
one, virtual-state solutions yield a very large radiative decay width, about
$500\div 800$ keV which, at present, has no reasonable explanation.
  
Furthermore, we have demonstrated that the new Belle data on the $D^0 \bar
D^{*0}$ mode are in conflict with the old Belle data
\cite{Belle3875}, as well as with the BaBar data on the same mode (up to
the resolution issue, as was mentioned before). If the Belle peak at $3872.6$
MeV is real, as suggested by the fine resolution and high statistics of the
new Belle data, then there is no need anymore to reconcile the $\pi^+\pi^-
J/\psi$ and $D \bar D^*$ peaks, and the best fit is consistent with 
the bound--state solution. However, to overcome the problem of the large
ratio (\ref{Sdpsi}) of the branching fractions, we are forced to include
``extra"
non-$D \bar D^*$ modes, with the radiative width $\varGamma(\gamma \psi')$ 
being only a small fraction of these ``extra" modes. 
Thus the difference between the present results and the ones of
Ref.~\cite{recon}
is due to the new data as well as the model-dependence allowed here.

The spectral density was calculated for all solutions presented and it appears
that, for the bound-state solutions, there is a significant admixture of the
bare state in the $X$ wave function. 
The properties of the bare state delivered by the bound-state solution are in
good agreement with the ones of the $\chi'_{c1}$ charmonium: the branching
fraction in the $B \to K$ decay, the total width, and the radiative
$\gamma \psi'$ width comply well with the charmonium assignment.
 
We stress that, in this picture, the $X$ is not a {\it bona fide} charmonium accidentally residing at the $D^0 \bar D^{*0}$ threshold. 
Had it been the case, the integral of the spectral density over the resonance
region would have been unity while, for our bound-state solutions, it does not
exceed 50\%. 
It is rather a
resonance attracted to the threshold, a phenomenon advocated in
Ref.~\cite{bugg2} and described in microscopical models in
Refs.~\cite{YuSK,DS}. In other words, the $X$ is generated 
dynamically by a strong coupling of the bare $\chi'_{c1}$ state to the $D \bar
D^*$ hadronic channel, with a large admixture of the $D \bar D^*$ molecular
component.

On the contrary, the virtual-state solution favoured by the BaBar data points
to a rather small (if any) admixture of the bare state in the $X$ wave function,
and there is no need to invoke ``extra" modes. This feature, in principle, could
discriminate between bound-state and virtual-state solutions.
In practice, the annihilation (light hadrons) modes encoded in the quantity
$\varGamma_0$ are not easily detectable so, in further studies, one is to rely
upon improvements in the data on already observed modes. 

In particular, a clear signature for a bound-state solution is the
below-threshold $D^0 \bar D^0 \pi^0$ peak.  
Unfortunately, from the experimental point of view, published data are not
decisive, mainly due to the kinematical cuts imposed by the assumption on
the $D^0
(\bar D^0) \pi^0$ mode coming from $D^{*0} (\bar D^{*0})$ one. In this regard,
we urge both experimental collaborations to overcome this and to perform an
unbiased analysis.

\begin{acknowledgments}
The authors are grateful to C. Hanhart for collaboration, reading the
manuscript, and critical comments, to T. Aushev (Belle Collaboration) and to
R. Faccini, M. A. Mazzoni, A. D'Orazio, and V. Poireau (BaBar Collaboration)
for valuable discussions of the experimental data.
This work was supported by the State Corporation of Russian
Federation ``Rosatom'' and by the grants RFFI-09-02-91342-NNIOa, DFG-436 RUS 
113/991/0-1(R), NSh-843.2006.2. A. N. would also like to acknowledge the
support of the grants RFFI-09-02-00629a, PTDC/FIS/70843/2006-Fi\-si\-ca, and of
the non-profit ``Dynasty'' foundation and ICFPM.
\end{acknowledgments}

\end{document}